\documentclass[a4paper,11pt]{article}
\pdfoutput=1 

\usepackage{jheppub} 

\usepackage[T1]{fontenc} 

\usepackage{graphicx}
\usepackage{comment}
\usepackage{subcaption}

\newcommand{\CI}{{\cal C}_1}
\newcommand{\CII}{{\cal C}_2}
\newcommand{\CIII}{{\cal C}_3}
\newcommand{\CIV}{{\cal C}_4}

\newcommand{\bea}{\begin{eqnarray}}
\newcommand{\eea}{\end{eqnarray}}
\def\no{\nonumber}

\newcommand{\p}{\partial}

\def\cB{{\cal B}}

\newcommand{\be}{\begin{equation}}
\newcommand{\ee}{\end{equation}}
\newcommand{\beq}{\begin{eqnarray}}
\newcommand{\eeq}{\end{eqnarray}}

\title{\boldmath Near Horizon Geometry of Strings Ending on Intersecting D8/D4-branes}

\author[a]{J. Estes,}
\author[b]{D. Krym,}
\author[c]{B. Van Pol,}

\affiliation[a]{Blackett Laboratory, Imperial College,\\London, SW7 2AZ, United Kingdom}
\affiliation[b]{New York City Ccollege of Technology \& Center for Theoretical Physics,\\300 Jay Street, Brooklyn, NY, 11228, USA}
\affiliation[c]{Instituut voor Theoretische Fysica, KU Leuven,\\ Celestijnenlaan 200D B-3001 Leuven, Belgium}

\emailAdd{johnaldonestes@gmail.com}
\emailAdd{daryakrym@gmail.com}
\emailAdd{bert.vanpol@fys.kuleuven.be}

\abstract{We consider solutions of massive IIA supergravity corresponding to the half-BPS intersection of D8/D4-branes with fundamental strings.  The $1+1$-dimensional intersection preserves the symmetry $D(2,1;\gamma;1) \times SO(4)$.  We give a reduction and partial integration of the BPS equations for this symmetry group.  We then specialize to the cases of enhanced supersymmetry corresponding to $\gamma = -1/2,-2$ or $\gamma = 1$.  In the first case, we show that the only solution with enhanced symmetry is given by the $AdS_6$ geometry describing the near horizon geometry of D8/D4-branes in the presence of an O8-plane.  In the second case, we identify novel solutions corresponding to fundamental strings ending on D8-branes and a second set of novel solutions corresponding to fundamental strings ending on an O8-plane.  In both cases, the fundamental string geometry contains an asymptotically flat region where the string coupling goes to zero.  We also show that there are no solutions corresponding to $1+0$-dimensional CFTs, which one may have hoped to construct by suspending fundamental strings between D8-branes.}

\preprint{Imperial/TP/2014/JE/02}

\begin{document}

\maketitle
\flushbottom

\section{Introduction}

\label{sec:intro}

The study of defects, interfaces or boundaries allows one to explore many interesting phenomena. A magnetic impurity in a material can be viewed as a defect, facilitating the description of the Kondo effect, which involves the screening of said impurity at low temperatures \cite{Kondo01071964}. If one puts a boundary in a system describing a superconductor, one may hope to describe surface superconductivity \cite{PhysRevLett.12.442}.  Topological insulators are constructed by joining two systems with different topological invariants along an interface \cite{PhysRevLett.49.405,2009AIPC.1134...22K,RevModPhys.82.3045}. A Wilson line operator, which measures the interaction energy or self-energy of charged particles, can also be viewed as a defect. There is a great deal of interest in understanding these types of effects in strongly coupled systems where standard perturbative approaches are no longer valid.

The AdS/CFT correspondence allows one to study strongly coupled systems, by instead, studying weakly coupled dual gravitational systems.  One may hope to understand the aforementioned phenomena in strongly coupled regimes by studying dual gravitational systems with defects, interfaces or boundaries.  Restricting to systems with large amounts of supersymmetry, large classes of geometries have been explicitly constructed.  These include geometries describing Wilson lines \cite{Lunin:2006xr,D'Hoker:2007fq}, surface operators \cite{Gomis:2007fi}, defect theories \cite{Bak:2003jk,Assel:2011xz} and theories with boundaries \cite{Aharony:2011yc}.  These explicit geometries have been used to provide non-trivial checks of the AdS/CFT correspondence \cite{Gomis:2006mv,Assel:2012cp}.  They have also been been used in holographic realizations of the Kondo effect \cite{Harrison:2011fs} and topological insulators \cite{Karch:2009sy}.

A particularly interesting AdS/CFT correspondence arises in the context of 5-dimensional gauge theories.  In general, 5-dimensional gauge theories are non-renormalizable and it is believed that additional degrees of freedom must be added to make the theories well defined.  One may even try to argue that such theories are never well defined and this is one of the reasons we live in 4-dimensions.  However, a proposed counter-example to this argument is provided by a class of 5-dimensional supersymmetric gauge theories which are conjectured to be UV-complete \cite{Seiberg:1996bd,Morrison:1996xf,Intriligator:1997pq}.  Evidence for the existence of a non-trivial UV fixed point is provided by the existence of a gravitational dual \cite{Ferrara:1998gv,Brandhuber:1999np,Bergman:2012kr}.  Evidence that the 5-dimensional gauge theory captures the UV fixed point physics has been provided by finding agreement between the $S^5$ partition function of the gauge theory and dual gravitational system \cite{Jafferis:2012iv}.  It is still an open question if the non-perturbative effects are sufficient to cure the UV-divergences or if additional degrees of freedom are necessary.

Recently, the partition function result has been extended to comparing the vacuum expectation value of the half-BPS Wilson line for totally symmetric and anti-symmetric representations \cite{Assel:2012nf}.  In either of these cases, the vacuum expectation value can be well approximated on the gravity side using probe branes.  To go further, one must include the backreaction of the probe branes.  In terms of branes, the background 5-dimensional gauge theory arises as the low energy limit of parallel D4-branes and D8-branes along with an O8 orientifold projection.  Introducing a Wilson line in the fundamental representation corresponds to introducing a fundamental string perpendicular to the D4/D8-brane system.  Rank $M$ Symmetric representations arise from introducing an additional D4-brane and stretching $M$ fundamental strings between the D4-brane and the D4/D8-brane stack.  Rank $M$ anti-symmetric representations arise by introducing a perpendicular D4-brane and $M$ fundamental strings.  The brane configurations are summarized in table \ref{tab:F1D4}.  In general, the BPS-Wilson line reduces the superconformal symmetry of the 5-dimensional gauge theory from $F(4;2) \times SU(2)$ to $D(2,1;2;1) \times SO(4)$\footnote{For supergroups, we follow the conventions of \cite{Frappat:1996pb}.}.

An analogous system exists in M-theory which consists of intersecting M2/M5-branes with supergroup $D(2,1;\gamma;1) \times D(2,1;\gamma;1)$.  In fact, the M-theory system is the M-theory uplift of the above configuration after removing the D8-branes and O8-plane.  In this case the 5-dimensional gauge theory of the D4-branes becomes a 6-dimensional CFT at the UV fixed point and the Wilson line operators in the gauge theory become surface operators in the 6-dimensional CFT.  The general M-theory solutions with $D(2,1;\gamma;1) \times D(2,1;\gamma;1)$ symmetry were constructed in \cite{D'Hoker:2008wc,Estes:2012vm}.  The specific solutions describing the 6-dimensional CFT with surface operator insertions were constructed in \cite{D'Hoker:2008qm}.

In addition to solutions describing the M5-brane solutions, additional solutions corresponding to the effective theory on the M2-branes were also found.  These correspond to the Janus solutions of \cite{D'Hoker:2009gg} and solutions which describe M2-branes ending on M5-branes of \cite{Bachas:2013vza}.  A natural question is whether there are analogous solutions after introducing the D8-branes and possibly the O8-plane.  Does the case of fundamental string ending on D8-branes admit a good decoupling limit?

With the task of finding backreacted geometries describing these Wilson lines in mind, we study general solutions of massive IIA supergravity with $D(2,1;\gamma;1) \times SO(4)$ symmetry.  We give a partial reduction and integration of the BPS equations, including obtaining algebraic expressions for the metric factors in terms of spinor bilinears.  These results are summarized in section \ref{sec:summBPS}.

In order to better understand the structure of the BPS system, we study the special cases of enhanced supersymmetry, which corresponds to setting certain fluxes to zero and setting $\gamma$ to specific values.  In general there are two distinct cases of enhanced supersymmetry, one given by setting $\gamma = -1/2,-2$ and the second given by setting $\gamma = 1$.  In the first case, we show the most general solution is given by the $AdS_6$ geometries of \cite{Brandhuber:1999np}, which is simply the dual of the 5-dimensional gauge theory without the half-BPS Wilson line.

The second case corresponds to fundamental strings ending on D8-branes.\footnote{We note this system was studied in \cite{Massar:1999sb} without the assumption of conformal symmetry.}  We identify three types of solutions.  The first, given in section \ref{sec:simpsol}, we interpret as a stack of fundamental strings in the presence of D8-branes, i.e. in a background with $F_0 \neq 0$\footnote{We denote the mass parameter of Roman's massive IIA supergravity by $F_0$.}.  The second two, given in section \ref{ssgensols}, we interpret as fundamental strings ending on a stack of D8-branes or an O8-plane.  In all three cases the geometry contains an asymptotically flat region.  Naively the geometry does not admit a decoupling limit.  However, we note that the string coupling goes to zero in the asymptotically flat regions, which may be sufficient for a valid decoupling limit.

We also consider solutions where $F_0$ is allowed to jump across an interface, corresponding to the presence of a stack of D8-branes.  This allows for a large family of solutions, parameterized by the number of such jumps.  However, we find that there is no way to glue D8-brane caps or O8-plane caps together.  Consequently, we argue that there are no solutions dual to $1+0$-dimensional CFTs.

In both cases, we identify a holomorphic $(1,0)$-form, which is a ubiquitous ingredient in half-BPS solutions.  In almost all known solutions, the holomorphic $(1,0)$-form is a homogenous polynomial of the spinor variables.  In the current case, the degree of the polynomial is different for the two distinct cases of enhanced symmetry.  This means that for the general solutions, the holomorphic $(1,0)$-form, if it exists, is not a polynomial of the spinor variables.  This suggests that the resulting BPS structure is different than the analogous M-theory system.

The paper is organized as follows.  In section \ref{sec:probe}, we give a further discussion of the brane configurations of interest in this paper.  In section \ref{sec:three}, we present our reduction of the BPS system for $D(2,1;\gamma;1) \times SO(4)$ invariant geometries, including a summary of the reduced equations.  In section \ref{sec:four}, we solve the BPS system for the special case $\gamma = -2$ with enhanced supersymmetry.  This corresponds to the case of D4/D8-branes.  In section \ref{sec:five}, we solve the BPS system for the special case $\gamma = 1$ with enhanced supersymmetry.  This corresponds to the case of fundamental strings ending on D8-branes or an O8-plane.

\section{Probe brane description}
\label{sec:probe}

In this section, we consider brane configurations which preserve $D(2,1;\gamma;1) \times SO(4)$ symmetry.  We first summarize results from an analogous system in M-theory with $D(2,1;\gamma;1) \times D(2,1;\gamma;1)$ symmetry, which corresponds to intersecting M2/M5-branes.  Compactifying to type IIA and introducing D8-branes and an O8 orientifold projection leads to the symmetry group $D(2,1;\gamma;1) \times SO(4)$.

We start by considering a related system in M-theory.  We consider intersecting M2-branes and M5-branes, with the configuration given in table \ref{tab:M2M5}. The intersection preserves 16 supersymmetries, where we count also the super conformal supersymmetries.  The bosonic symmetry preserved by the brane configuration in the conformal limit is given by $SO(2,2) \times SO(4) \times SO(4)$.  The corresponding superconformal group with $SO(2,2) \times SO(4) \times SO(4)$ bosonic symmetry and 16 supersymmetires is not unique but comes in a one-parameter family, $D(2,1;\gamma;1) \times D(2,1;\gamma;1)$.  In \cite{D'Hoker:2008wc}, the problem of finding solutions of M-theory with $D(2,1;\gamma;1) \times D(2,1;\gamma;1)$ symmetry was reduced to a single linear partial differential equation.  At three special points, the $D(2,1;\gamma;1)$ supergroup reduces to a classical supergroup
\begin{align}
\label{specgroup}
&D(2,1;\gamma;1) = OSp(4^*|2)& &\gamma = -1/2, -2\,,& \cr
&D(2,1;\gamma;1) = OSp(4|2,\mathbf{R})& &\gamma = 1\,.&
\end{align}
Furthermore, at each of these three special points, the superconformal group becomes a subgroup of a larger superconformal group.  This group structure leads to a large and interesting family of solutions.
\begin{table}[h]
\begin{center}
\begin{tabular}{|c||c|c|c|c|c|c|c|c|c|c|c|}
  \hline
      & 0 & 1 & 2 & 3 & 4 & 5 & 6 & 7 & 8 & 9 & 10 \\
      \hline
  M2   & X & X &   &   &   &   &   &   &   &   &  X \\
  M5   & X &   & X & X & X & X &   &   &   &   &  X \\
  M5'  & X &   &   &   &   &   & X & X & X & X &  X \\
  \hline
\end{tabular}
\end{center}
\caption{Intersecting M2/M5-branes, preserving half of the maximal number of supersymmetries.}
\label{tab:M2M5}
\end{table}

We consider first the special values $\gamma = -1/2,-2$.  In this case we have $D(2,1;-1/2,0) \times D(2,1;-1/2;0) \subset OSp(8^*|4)$ and  $D(2,1;-2;0) \times D(2,1;-2;0) \subset OSp(8^*|4)$.  The extended symmetry $OSp(8^*|4)$ corresponds to having only a single stack of M5-branes and is realized as the superisometry group of the near horizon geometry $AdS_7 \times S^4$.  It is conjectured that the dual field theory is given by the 6d $(2,0)$ SCFT with $OSp(8^*|4)$ symmetry, although no completely satisfactory construction of this theory exists.  It is believed that this theory admits supersymmetric surface operators (also known as the self-dual string), which preserve half of the supersymmetries and break the symmetry to $OSp(4^*|2) \oplus OSp(4^*|2)$.\footnote{The surface operators can be thought as a generalization of Wilson loops.} These operators arise from introducing the M2-branes of table \ref{tab:M2M5}.  A single M2-brane produces a surface operator in the fundamental representation of the gauge group, while multiple M2-branes produce a surface operator in a higher rank representation.  Dual supergravity solutions describing the surface operators have been constructed in \cite{D'Hoker:2008qm}.

For the special value $\gamma = 1$, the extended symmetry is $OSp(8|4,\mathbf{R})$ and corresponds to having a stack of M2-branes, whose near horizon geometry is $AdS_4 \times S^7$.  The dual field theory is given by a 3d ${\cal N} = 8$ SCFT.  A Lagrangian description of this theory is given by ABJM theory, although only ${\cal N} = 6$ supersymmetry is manifest in the Lagrangian.  Consider now the brane configuration of table \ref{tab:M2M5}.  Naively, the intersecting M5-branes gives rise to a 2d defect preserving half of the supersymmetries, although such a superconformal defect has yet to be constructed either directly in the field theory or in the dual supergravity solutions.

Instead of inserting the M5-branes as a defect, we can also consider the case where the M2-branes end on the M5-branes.  In the dual field theory, this corresponds to putting ABJM theory on a half-space with supersymmetric boundary conditions, possibly coupled to 2d degrees of freedom localized on the boundary.  Such supersymmetric boundary conditions were studied in \cite{Berman:2009kj,Faizal:2011cd,Okazaki:2013kaa} and dual supergravity solutions were constructed in \cite{Bachas:2013vza}, although a precise holographic dictionary has yet to be constructed.  Since the M2-branes end on the M5-branes, the supersymmetry does not necessarily need to be a subgroup of $OSp(8|4,\mathbf{R})$, as there is no way to remove the M5-branes and recover the full M2-brane theory.  Indeed the solutions of \cite{Bachas:2013vza} have $D(2,1;\gamma,0) \times D(2,1;\gamma,0)$ symmetry with $\gamma < 0$.

The above story can be reduced to type IIA string theory after compactifying and dimensionally reducing along the $10$-direction.  The resulting type IIA brane configuration is given in the top half of table \ref{tab:F1D4}.  The M2-branes become a fundamental string, while the M5-branes become D4-branes.
\begin{table}
\begin{center}
\begin{tabular}{|c||c|c|c|c|c|c|c|c|c|c|}
  \hline
       & 0 & 1 & 2 & 3 & 4 & 5 & 6 & 7 & 8 & 9  \\
      \hline
  F1   & X & X &   &   &   &   &   &   &   &     \\
  D4   & X &   & X & X & X & X &   &   &   &     \\
  D4'  & X &   &   &   &   &   & X & X & X & X   \\
   \hline
  O8/D8   & X &   & X & X & X & X & X & X & X & X   \\
  D0   & X &   &   &   &   &   &   &   &   &     \\
  \hline
\end{tabular}
\end{center}
\caption{Intersecting D4-branes and F1-strings, preserving half of the maximal number of supersymmetries.  }
\label{tab:F1D4}
\end{table}

We now move on to the system of interest in this paper.  Starting with the above IIA configuration, we introduce D8-branes, along with an orientifold projection, as in the lower half of table \ref{tab:F1D4}.\footnote{The symmetry projections of the branes given in table \ref{tab:F1D4} are given by
\begin{align}
&\rm{F1-string}& &\epsilon = \Gamma_{23456789} \epsilon\,,&
&\rm{D4-brane}& &\epsilon = \Gamma_{16789} \epsilon\,,& \cr
&\rm{D4'-brane}& &\epsilon = \Gamma_{12345} \epsilon\,,&
&\rm{D8-brane}& &\epsilon = \Gamma_{1} \epsilon\,,& \cr
&\rm{D0-brane}& &\epsilon = \Gamma_{123456789} \epsilon\,.& \no
&&&& \no
\end{align}
The fundamental string projection breaks the number of supersymmetries from $32 \rightarrow 16$, while the D4-brane projection further breaks the number of supersymmetries to $8$.  Introducing the D8-branes further reduces the supersymmetries by half, so that the resulting theory preserves $4$ Poincar\'e supersymmetries.  In the conformal limit, the total number of supersymmetries doubles.}  The D8-branes/O8-plane reduce the supersymmetry further and only 8 supersymmetries remain.  The D8-branes source the Roman's mass and we work in massive IIA supergravity.  Note that we may also introduce D0-branes without further reducing the symmetry.  The bosonic symmetry in the conformal limit is given by $SO(1,2) \times SO(4) \times SO(4)$.  The corresponding superconformal group again comes in a one parameter family, $D(2,1;\gamma;1) \times SO(4)$, where we have an extra bosonic symmetry which is not part of the supergroup.  As in M-theory, for three special values of $\gamma$, we find that $D(2,1;\gamma;1) \times SO(4)$ is a subgroup of an extended supergroup:
\begin{align}
&D(2,1;\gamma;1) \times SO(4) \subset F(4;2) \times SO(3)& &\gamma = -1/2, -2\,,& \cr
&D(2,1;\gamma;1) \times SO(4) \subset OSp(8|2,\mathbf{R})& &\gamma = 1\,.&
\end{align}
We see a group structure analogous to the one encountered for M-theory and it is a natural question to ask what the enhanced symmetry points correspond to.

For the special values $\gamma = -1/2,-2$, the extended symmetry $F(4;2) \times SO(3)$ corresponds to a single stack of D4-branes in addition to the D8-branes/O8-plane.  The low energy field theory consists of a 5-dimensional ${\cal N} =2$ gauge theory with gauge group $USp(2N)$, a single antisymmetric hypermultiplet and $N_f$ fundamental hypermultiplets, where there are $2N$ D4-branes and $N_f$ D8-branes.\footnote{The antisymmetric hypermultiplet arise from D4-D4 strings stretched across the O8-plane, while the $N_f$ fundamental hypermultiples come from D4-D8 strings.}  It is believed that this theory is ultraviolet complete with a 5d ${\cal N}=2$ ultraviolet superconformal fixed point.  Field theory arguments for this conjecture are given in \cite{Seiberg:1996bd,Morrison:1996xf,Intriligator:1997pq}.  Dual solutions of massive IIA supergravity have been constructed in \cite{Ferrara:1998gv,Brandhuber:1999np}, which provide further evidence for the existence of the ultraviolet fixed point.

We now consider what happens when we introduce the fundamental strings.  Introducing the fundamental string reduces the supersymmetry by half and in the field theory corresponds to introducing a half-BPS Wilson line.  The superconformal symmetry is broken to $OSp(4^*|2) \subset F(4;2)$, while the $SO(3)$ bosonic symmetry remains unbroken, since the Wilson line is neutral.  To see this, note that a time-like Wilson line preserves an $SO(4) = SO(3) \times SO(3)$ rotational symmetry and translations in time, which in the conformal limit are enhanced to $SO(1,2)$.  This gives a full bosonic symmetry $SO(1,2) \times SO(4) \times SO(4)$, which can be seen in table \ref{tab:F1D4}.  The dual supergravity solutions have yet to be constructed but fall within the anstaz studied in this paper.

The special value of $\gamma = 1$ corresponds to the extended symmetry $OSp(8|2,\mathbf{R})$.  This symmetry arises as the superconformal symmetry associated with fundamental strings ending on or intersecting the D8-branes/O8-plane.  This can be seen in table \ref{tab:F1D4}, where the $SO(8)$ symmetry acts on the directions $2$ through $8$, while $Sp(2,\mathbf{R}) \sim SO(1,2)$ arises as the superconformal group of time translations.  The putative dual field theory would correspond to a 2-dimensional boundary or defect CFT and would be the massive IIA analogue of the M2-brane near horizon geometry.  Additionally, we may include the D0-branes without loss of symmetry.  In this case, the fundamental strings can be stretched between the D8-branes and the D0-branes.

\section{$D(2,1;\gamma;1) \times SO(4)$ invariant geometries}
\label{sec:three}

We consider first the most general solutions of massive IIA supergravity which preserve the symmetry $D(2,1;\gamma;1) \times SO(4)$.  The bosonic subalgebra $so(1,2) \times so(4) \times so(4)$ is naturally realized on the $10$-dimensional space-time $AdS_2 \times S^3 \times S^3 \times \Sigma_2$, where $\Sigma_2$ is a 2-dimensional space.  The metric takes the form
\begin{align}
ds^2 = f_1^2 ds^2_{AdS_2} + f_2^2 ds^2_{S^3} + f_3^2 ds^2_{S^3} + ds_{\Sigma_2}^2.
\end{align}
It will be convenient to introduce the frames
\begin{align}
e^m& = f_1 \hat e^m& m=&0,1, \qquad&
e^{i}& = f_2 \hat e^{i}& i=&2,3,4, \qquad& \cr
e^{\tilde i}& = f_3 \hat e^{\tilde i}& \tilde i=&5,6,7, \qquad&
e^a & & a=&8,9, \qquad&
\end{align}
where $\hat e^m$ are frames on the unit $AdS_2$, $\hat e^{i}$ and $\hat e^{\tilde i}$ are frames on the two unit $S^3$'s and $e^a$ are frames on $\Sigma_2$.
The dilaton $\phi$ and the warp factors $f_i$ are restricted to be functions of $\Sigma_2$.
The 4-form flux takes the form
\begin{align}
F_4 = h_1 e^{0189} + h_a e^{234a} + g_a e^{567a}
\end{align}
while the 2-form gauge potential and corresponding 3-form field strength take the form
\begin{align}
B_2 =& b_0 \hat e^{01} +  b_1 e^{89},&
H_3 =& d B_2 = d b_0 \wedge \hat e^{01}.&
\end{align}
The coefficients $h_1$, $h_a$, $g_a$, $b_0$ and $b_1$ are all functions of $\Sigma_2$.
The Romans mass \cite{Romans:1985tz} is denoted $F_0$.

In order to preserve the supersymmetries, the supersymmetry variations of the fermionic fields must vanish.  From \cite{Romans:1985tz} we get the following BPS equations modulo some redefinitions, spelled out in appendix \ref{app:typeIIA}
\begin{align}
\label{BPSeqns}
0 =& \bigg[ (D_M \phi) \Gamma^M + \frac{5}{4} F_0
 e^{\frac{5}{4} \phi} + \frac{1}{96} e^\frac{\phi}{4} (F_{MNPQ}
 \Gamma^{MNPQ})
 \cr &\qquad
 - \frac{3}{8} F_0  e^{\frac{3\phi}{4}} B_{MN} \Gamma^{MN} \Gamma_{11}
 - \frac{1}{12} e^{-\frac{\phi}{2}} H_{MNP} \Gamma^{MNP} \Gamma_{11}
 \bigg] \epsilon  \ , \no
\end{align}
\begin{align}
0 =& \bigg[ D_M - \frac{1}{32} F_0 e^{\frac{5}{4} \phi}
 \Gamma_M + \frac{1}{128} \frac{e^{\frac{\phi}{4}}}{2} F_{NPQR} (
 \Gamma_M {}^{NPQR} - \frac{20}{3} \delta_M {}^N \Gamma^{PQR} )
 \cr&\qquad
 - \frac{1}{32} F_0 \frac{e^{\frac{3\phi}{4}}}{2} B_{NP} (\Gamma_M {}^{NP} - 14 \delta_M {}^N \Gamma^P) \Gamma_{11}
  \cr&\qquad
 + \frac{1}{48} \frac{e^{-\frac{\phi}{2}}}{2} H_{NPQ} (\Gamma_M {}^{NPQ} - 9 \delta_M {}^N \Gamma^{PQ}) \Gamma_{11}
 \bigg]
 \epsilon \ .
\end{align}
Our conventions for $\Gamma$-matrices are defined in appendix \ref{app:gamma}.

Since the $B_2$ field has a mass in massive IIA supergravity, the theory is no longer invariant under gauge transformations of $B_2$.  In \cite{Romans:1985tz}, this fact has been used to absorb the Ramond-Ramond $2$-form, $F_2$ into the definition of $B_2$.  In order to connect back to the massless IIA supergravity, we make the replacement $B_{2} \rightarrow B_{2} - F_0^{-1} F_{2}$ and $C_{3} \rightarrow C_{3} - (2 F_0)^{-1} A_{1} \wedge F_{2}$ with $d F_{2} = 0$  and $F_{2} = d A_{1}$.  The theory then enjoys the symmetry
\begin{align}
\label{MIIAGT}
&B_{2} \rightarrow B_{2} + d \Lambda\,,&
&F_{2} \rightarrow F_{2} + F_0 \, d \Lambda\,,&
&C_{3} \rightarrow 2 \Lambda \wedge F_2 + 2 C_1 \wedge d \Lambda + 2 F_0 \, \Lambda \wedge d \Lambda\,,&
\end{align}
where $\Lambda$ is a $1$-form.  To obtain massless IIA supergravity, we simply take $F_0=0$.

The fluxes of our ansatz have the following interpretation in terms of the brane constructions discussed in section \ref{sec:probe}.  The $h_a$ and $g_a$ components of $F_4$ correspond to the D4-brane and D4'-branes of table \ref{tab:F1D4}.  In massless type IIA theory, the fundamental string sources the NS-NS three form $H_3$, while the D0-branes source the Ramond-Ramond two form $F_2$. In massive IIA supergravity, the gauge transformation \eqref{MIIAGT} mixes them.  A related effect occurs in the brane description.  When one pulls a D0-brane through a D8-brane, a fundamental string is created which stretches between the D8-brane and the D0-brane \cite{Danielsson:1997wq}.  The D8-branes can be interpreted as a magnetic source for the scalar field strength $F_0$. Since we are interested in the brane configurations of section \ref{sec:probe} and the D4 brane sources the $F_4$ field strength only magnetically, we set $h_1$ to zero. Similarly, the fundamental string sources the $B_2$ gauge potential only electrically so we set $b_1$ to zero.

\subsection{BPS equations}
\label{sec:BPSeqns}

Our first step is to reduce the BPS equations \eqref{BPSeqns} to a two-dimensional system.  This is carried out as follows.  First the supersymmetry parameter, $\epsilon$, is decomposed using a basis of Killing spinors for the symmetric spaces $AdS_2$ and $S^3$.  We denote by $\chi_{\eta_1,\eta_2,\eta_3}$ a basis of Killing spinors on $AdS_2 \times S^3 \times S^3$, where $\eta_i = \pm 1$.  These can be explicitly constructed following \cite{D'Hoker:2007xy} and satisfy the Killing spinor equations \eqref{eq:KS}.  The basis, $\chi_{\eta_1,\eta_2,\eta_3}$, is actually overcomplete so without loss of generality, we impose the conditions $\chi_{\eta_1,\eta_2,\eta_3} = \left( B_{(1)} \otimes B_{(2)} \otimes B_{(3)} \right) \chi_{\eta_1,\eta_2,\eta_3}^*$ and $\chi_{\eta_1,\eta_2,\eta_3} = (\sigma^3 \otimes I_2 \otimes I_2) \chi_{-\eta_1,\eta_2,\eta_3}$.  Note that these conditions are consistent with the Killing spinor equations \eqref{eq:KS}.  We then write the 10-dimensional supersymmetry parameter as
\begin{align}
\epsilon = \sum_{\eta_1,\eta_2,\eta_3} \chi_{\eta_1,\eta_2,\eta_3} \otimes \left[ \zeta_{\eta_1,\eta_3,\eta_3} \otimes \left(
                                                                                                                  \begin{array}{c}
                                                                                                                    1 \\
                                                                                                                    0 \\
                                                                                                                  \end{array}
                                                                                                                \right)
+  \hat \zeta_{\eta_1,\eta_3,\eta_3} \otimes \left(
                                                                                                                  \begin{array}{c}
                                                                                                                    0 \\
                                                                                                                    1 \\
                                                                                                                  \end{array}
                                                                                                                \right)
\right]\,,
\end{align}
where the coefficients of the decomposition, $\zeta_{\eta_1,\eta_3,\eta_3}$ and $\hat \zeta_{\eta_1,\eta_3,\eta_3}$, are collections of two-component spinors.  The type IIA reality condition $\epsilon^* = {\cal B} \epsilon$ relates the coefficients $\zeta$ and $\hat \zeta$ by $\hat \zeta^* = i \sigma_2 \zeta$.  The BPS equations can then be written as equations on $\zeta$, the coefficients of the decomposition.  This is carried out explicitly for the first equation of \eqref{BPSeqns} in appendix \ref{app:red1}.  The final result is summarized in equations \eqref{app1:dil}--\eqref{app1:grav5}.  In giving these expressions, we have written $\zeta$ as a single $16$-component spinor with the $\eta_i$-labels corresponding to spin indices.  We also introduced the notation $\tau^{(ijk)} = \sigma^i \otimes \sigma^j \otimes \sigma^k$, where the $i$-th Pauli matrix acts on the $\eta_i$ index.

To simplify the problem, we first look for symmetries of the equations \eqref{app1:dil}--\eqref{app1:grav5}.  Both $\tau^{(030)}$ and $\tau^{(003)}$ commute with the BPS equations.  In the special case $b_1 = h_1 = 0$, $\tau^{(300)} \sigma^3$ also commutes with the BPS equations.  We are interested in the brane configurations discussed in section \ref{sec:probe} and correspondingly set $b_1 = h_1 = 0$ throughout the rest of the paper.  The reduced BPS equations for general values of $b_1$ and $h_1$ are given in appendix \ref{app:gencase}.
As a result of the three symmetries, we can impose the following projections on $\zeta$ without loss of generality
\begin{align}
\label{eq:proj}
\zeta =& \nu_1 \tau^{(300)} \sigma^3 \zeta\,,&
\zeta =& \nu_2 \tau^{(030)} \zeta\,,&
\zeta =& \nu_3 \tau^{(003)} \zeta\,,&
\end{align}
where the $\nu_i$ are each a sign choice for the projection.  The last two projections simply project onto the components of $\zeta_{\eta_1,\eta_2,\eta_3}$ with $\eta_2 = \nu_2$ and $\eta_3 = \nu_3$.  The first projection is solved for the case $\nu_1 = +1$ by setting $\zeta_{-+} = \zeta_{+-} = 0$, where the first index corresponds to $\eta_1$ and the second to the spinor index.  In this case, we may group the remaining components into a variable $\xi = (\zeta_{++},\zeta_{--})$.  In the case $\nu_1 = -1$, we have $\zeta_{++} = \zeta_{--} = 0$ and we group the remaining components into a variable $\xi = (\zeta_{-+},\zeta_{+-})$.

After imposing the above projections, the BPS equation \eqref{app1:dil} reduces to
\begin{align}
\label{eq:odil}
0&=\frac{1}{2\sqrt{2}}D_z \phi \sigma^3 \xi^* + \frac{5}{8\sqrt{2}}F_{0}e^{5\phi/4} i \sigma^2 \xi -\frac{1}{8\sqrt{2}}e^{\phi/4}(h_z \sigma^2 \xi^*  + g_z \sigma^1 \xi^* ) \nonumber\\
&+ \frac{3}{8\sqrt{2}}F_{0}e^{3\phi/4} \frac{b_0}{f_1^2} \sigma^1 \xi - \frac{1}{4\sqrt{2}}e^{-\phi/2} \frac{D_{z}b_0}{f_1^2} \xi^*\,.
\end{align}
The equations \eqref{app1:grav1}--\eqref{app1:grav3} reduce to
\begin{align}
\label{eq:oalg}
0=& -\frac{\nu_1}{2f_1} \sigma^3 \xi+\frac{1}{2}D_{z}\ln f_{1} \sigma^1 \xi^* - \frac{1}{32}F_{0}e^{5\phi/4}\xi + \frac{3}{16}\frac{e^{\phi/4}}{2}\left(i h_{z} \xi^* + g_z \sigma^3 \xi^* \right)
\nonumber\\ &
+ \frac{7}{16}F_{0}\frac{e^{3\phi/4}}{2} \frac{b_{0}}{f_{1}^{2}}\sigma^3\xi
+\frac{3}{8}\frac{e^{-\phi/2}}{2}\frac{D_z b_0}{f_1^2} i \sigma^2 \xi^*\,,
\cr
0=&
-\frac{i \nu_2}{2f_2}\sigma^{1} \xi +\frac{1}{2}D_{z}\ln f_{2} \sigma^1 \xi^* - \frac{1}{32}F_{0}e^{5\phi/4}\xi
+\frac{3}{16}\frac{e^{\phi/4}}{2}\left(g_z \sigma^3 \xi^*-i \frac{5}{3}h_{z} \xi^* \right)
\nonumber\\ &
-\frac{1}{16}F_{0}\frac{e^{3\phi/4}}{2} \frac{b_{0}}{f_{1}^{2}}\sigma^3\xi
-\frac{1}{8}\frac{e^{-\phi/2}}{2}\frac{D_z b_0}{f_1^2}  i\sigma^2 \xi^*\,, \cr
0=&
-\frac{i \nu_3}{2f_3}\sigma^{2} \xi+\frac{1}{2}D_{z}\ln f_{3} \sigma^1 \xi^* - \frac{1}{32}F_{0}e^{5\phi/4}\xi +\frac{3}{16}\frac{e^{\phi/4}}{2}\left(i h_z \xi^* -\frac{5}{3}g_{z} \sigma^3 \xi^*\right)
\nonumber\\&
-\frac{1}{16}F_{0}\frac{e^{3\phi/4}}{2} \frac{b_{0}}{f_{1}^{2}}\sigma^3\xi
-\frac{1}{8}\frac{e^{-\phi/2}}{2}\frac{D_z b_0}{f_1^2} i \sigma^2 \xi^*\,.
\end{align}
Note that these equations are algebraic in $\xi$.  The remaining equations \eqref{app1:grav4}--\eqref{app1:grav5} reduce to
\begin{align}
\label{eq:odiff}
0=&D_{z}\xi - \frac{1}{2} D_z(\ln \rho) \,  \xi
+\frac{3}{16}\frac{e^{\phi/4}}{2}\left[i \frac{8}{3} h_z \sigma^1 \xi + i \frac{8}{3} g_z \sigma^2  \xi \right]
-\frac{1}{2}\frac{e^{-\phi/2}}{2}\frac{D_{z} b_0}{f_1^2} \sigma^3 \xi \,,\cr
0=&D_{\bar z} \xi + \frac{1}{2} D_{\bar z} (\ln \rho) \xi - \frac{1}{16}F_{0}e^{5\phi/4} \sigma^1 \xi^*
+\frac{3}{16}\frac{e^{\phi/4}}{2}\left[i \frac{2}{3} h_{\bar z} \sigma^1 \xi + i \frac{2}{3} g_{\bar z} \sigma^2 \xi \right]\nonumber\\&
+i \frac{1}{8}F_{0}\frac{e^{3\phi/4}}{2} \frac{b_{0}}{f_{1}^{2}} \sigma^2 \xi^*
-\frac{1}{4}\frac{e^{-\phi/2}}{2}\frac{D_{\bar z} b_0}{f_1^2} \sigma^3 \xi\,,
\end{align}
which are differential in $\xi$.

We now examine the number of supersymmetries present in the system.  Each $\chi_{\eta_1,\eta_2,\eta_3}$ has 8 parameters and leads to 8 supersymmetries.  The projections \eqref{eq:proj} project onto specific values for $\eta_2$ and $\eta_3$ so that $\eta_2 = \nu_2$ and $\eta_3 = \nu_3$, while the components with different values of $\eta_1$ are related by the constraint $\chi_{\eta_1,\eta_2,\eta_3} = (\sigma^3 \otimes I_2 \otimes I_2) \chi_{-\eta_1,\eta_2,\eta_3}$.  Taking the background fields fixed, we will have 8 supersymmetries for each independent solution of $\xi$ to the above equations.  Generally we expect only one such solution.  To look for cases with enhanced supersymmetry, we first look for symmetries of the equations and consider sending $\xi \rightarrow M \xi$.  Requiring $M$ to commute or anti-commute with the first two terms of \eqref{eq:odil} leads to the possibilities $M = \sigma^1$, $i \sigma^2$ and $i \sigma^3$.  As a result we find the following three symmetires
\begin{align}
\label{eq:doublesusy}
&\xi \rightarrow \sigma^1 \xi&
&b_0 \rightarrow -b_0&
&g_z \rightarrow - g_z&
&\nu_{1(3)} \rightarrow - \nu_{1(3)}\,,& \cr
&\xi \rightarrow i \sigma^2 \xi&
&b_0 \rightarrow -b_0&
&h_z \rightarrow - h_z&
&\nu_{1(2)} \rightarrow - \nu_{1(2)}\,,& \cr
&\xi \rightarrow i \sigma^3 \xi&
&g_z \rightarrow - g_z&
&h_z \rightarrow - h_z&
&\nu_{2(3)} \rightarrow - \nu_{2(3)}\,.&
\end{align}
Note that in general these are not symmetries of the background fields, since we are required to flip the signs of the fluxes.  However, we can see that whenever two of the fluxes vanish, we will double the number of supersymmetries.  This is in agreement with the brane discussion of section \ref{sec:probe}.

In appendix \ref{app:red2}, we carry out a further reduction of the equations.  We first integrate the equations of \eqref{eq:oalg} which are differential in the $f_i$.  We obtain
\begin{align}
\label{eq:met}
&f_1 = \frac{\nu_1}{c_1} \xi^\dagger \xi\,,&
&f_2 = \frac{\nu_2}{c_2} \xi^\dagger \sigma^2 \xi\,,&
&f_3 = \frac{\nu_3}{c_3} \xi^\dagger \sigma^1 \xi\,,&
\end{align}
where the $\nu_i$ are sign choices and the $c_i$ are constants.  Using the remaining equations of \eqref{eq:oalg}, we obtain an algebraic constraint for the $c_i$
\begin{align}
c_1 - 2 c_2 + 2 c_3 = 0\,,
\end{align}
and an expression for $b_0$
\begin{align}
b_0 = \frac{8\xi^\dagger \xi}{F_{0} c_1^2}e^{-3\phi/4}\left(\frac{1}{2} c_1 - \frac{3}{4} c_2 + \frac{3}{4} c_3
+ \frac{1}{8}F_{0}e^{5\phi/4} \xi^\dagger \sigma^3 \xi\right)\,.
\end{align}
We introduce the notation $b_z = D_z b_0/f_1^2$ and treat $b_z$ as an independent variable from $b_0$.  We show that the BPS equations correctly enforce the differential relation between them.  The final summary of reduced equations is given in section \ref{sec:summBPS}.  There we denote the components of $\xi$ by $\alpha$ and $\beta$.

\subsection{Equations of motion and Bianchi identities}

In general the BPS equations are not sufficient to determine a valid supergravity solution.  Namely, there can be additional constraints arising from the Bianchi identities and equations of motion.  With this in mind, we first look at the Bianchi identities.  Since we give an anstaz directly for $B_2$, the Bianchi identity for $H_3$ is automatic.  The Bianchi identity for $F_4$ takes the form
\begin{align}
\label{eq:Bianchi}
d F_4 = 0
\qquad
\Rightarrow
\qquad
\left\{
\begin{array}{lc}
\p_z \left( f_2^3 \rho h_{\bar z} \right) - \p_{\bar z} \left( f_2^3 \rho h_z \right) = 0\,, \\
\p_z \left( f_3^3 \rho g_{\bar z} \right) - \p_{\bar z} \left( f_3^3 \rho g_z \right) = 0\,.
\end{array} \right.
\end{align}
We solve these equations by introducing the real functions $\varphi_{1}$ and $\varphi_{2}$ as $\p_{\bar z} \varphi_1 = \nu_2 f_2^3 \rho h_{\bar z}$ and $\p_{\bar z} \varphi_2 = \nu_3 f_3^3 \rho g_{\bar z}$.  The Bianchi identities then become integrability conditions for the fields $\varphi_i$.

Next we check that the Bianchi identities and BPS equations imply the equations of motion.  In \cite{Lust:2004ig}, it was shown that supersymmetry, together with the Maxwell equations and Bianchi identities imply that the dilaton equation and most of the Einstein equations are automatically satisfied.  More specifically, if we denote the Einstein equations collectively by $E_{MN}$, one finds that $E_{MN} = 0$ provided $E_{0M} = 0$ for $M \neq 0$.  Since our solution has an $AdS_2$ isometry, we have $E_{mn} \propto \eta_{mn}$ and $E_{mi} = E_{m\tilde i} = E_{ma} = 0$ and the condition is automatic.  As a result, we need only to check the Maxwell equations.

The Maxwell equations are given by
\begin{align}
0 =&\nabla_P \left( e^{-\phi} H^{PMN} \right) - F_{0}^2 e^{3\phi/2} B^{MN} - F_{0} e^{\phi/4} \frac{1}{2} F^{MNPQ} B_{PQ}
\no\\&
+ \frac{1}{24^2} \frac{1}{2} \epsilon^{MNPQRSTUVW} F_{PQRS} F_{TUVW}\,, \no\\
0 =&\nabla_Q \left( e^{\phi/2} F^{QMNP} \right) + \frac{1}{72} \frac{1}{2} \epsilon^{MNPQRSTUVW} F_{QRST} H_{UVW}\,.
\end{align}
Using our ansatz, these equations reduce to a set of three equations\footnote{We are using the notation $b_z = D_z b_0/f_1^2$.}
\bea
\label{maxeq}
0&=& D_{z}b_{\bar{z}}+D_{\bar{z}}b_{z}- D_z\phi b_{\bar{z}} - D_{\bar{z}}\phi b_z + i h_{\bar{z}}g_{z} e^{\phi} -i h_z g_{\bar{z}}e^{\phi} \nonumber\\ & &  \qquad
+b_z D_{\bar{z}}\ln\left(f_2^3 f_3^3 \rho\right) + b_{\bar{z}} D_{z}\ln\left(f_2^3 f_3^3 \rho\right) - 2 F_{0}^2 \frac{b_0}{f_1^2}e^{5\phi/2}\,,  \nonumber\\
0&=&D_{z}h_{\bar{z}}+D_{\bar{z}}h_{z}+\frac{1}{2}D_z\phi h_{\bar{z}} + \frac{1}{2}D_{\bar{z}}\phi h_z +i b_{\bar{z}} g_z e^{-\phi/2} -i b_z g_{\bar{z}}e^{-\phi/2} \nonumber\\&& \qquad
+ h_z D_{\bar{z}}\ln\left(f_1^2 f_3^3 \rho\right) + h_{\bar{z}} D_{z}\ln\left(f_1^2 f_3^3 \rho\right)\,,  \nonumber\\
0&=&D_{z}g_{\bar{z}}+D_{\bar{z}}g_{z}+\frac{1}{2}D_z\phi g_{\bar{z}} + \frac{1}{2}D_{\bar{z}}\phi g_z -i b_{\bar{z}} h_z e^{-\phi/2}  +ib_z h_{\bar{z}}e^{-\phi/2} \nonumber\\&& \qquad
+ g_z D_{\bar{z}}\ln\left(f_1^2 f_2^3 \rho\right)
+ g_{\bar{z}} D_{z}\ln\left(f_1^2 f_2^3 \rho\right)\,.
\eea
To check these equations, we employ the following strategy.  We use the algebraic equations to eliminate $g_z$, $b_0$ and $b_z$ in terms of $h_z$, $\alpha$, $\beta$ and $\phi$.  The Bianchi identities, \eqref{eq:Bianchi}, are then used to obtain expressions for $\p_z h_{\bar z}$ and its complex conjugate.  Along with the BPS equations, \eqref{eq:odil} and \eqref{eq:odiff}, this allows us to eliminate all derivative terms appearing in \eqref{maxeq}.  The computations are straight forward but tedious and we do not present them here.  The net result, once all derivatives have been eliminated is that the Maxwell equations are all automatically satisfied.  Thus in our case, it is sufficient to keep only the BPS equations and Bianchi identities.

\subsection{Summary of equations}
\label{sec:summBPS}

We give a summary of the remaining BPS equations and Bianchi identities.  The constraint for the $c_i$ and the Bianchi identities are
\begin{align}
\label{sum:e1}
c_1 - 2 c_2 + 2 c_3 = 0\,, \qquad
h_{z} = \frac{\nu_2}{f_2^3 \rho} \p_{z} \varphi_1\,, \qquad
g_{z} = \frac{\nu_3}{f_3^3 \rho} \p_{z} \varphi_2\,.
\end{align}
The metric factors are given by
\begin{align}
\label{sum:metfacts}
f_1 =& \frac{\nu_1}{c_1} (|\alpha|^2 + |\beta|^2)\,,&
f_2 =& \frac{i \nu_2}{c_2} (\alpha \beta^* - \alpha^* \beta)\,,&
f_3 =& \frac{\nu_3}{c_3} (\alpha \beta^* + \alpha^* \beta)\,.&
\end{align}
where $\alpha$ and $\beta$ are spinor components.
There are three equations algebraic in $\alpha$ and $\beta$
\begin{align}
\label{algconsts}
0 =& - \frac{F_0}{8}e^{3\phi/4} \frac{b_0}{f_1^2}  \left( |\alpha|^2 + |\beta|^2 \right)
+ \frac{1}{2} c_1 - \frac{3}{4} c_2 + \frac{3}{4} c_3
+ \frac{1}{8}F_0e^{5\phi/4} \left( |\alpha|^2 - |\beta|^2 \right), \cr
0=&2\left(c_2 + c_3 \right)
+ e^{\phi/4} g_z \left( (\alpha^*)^2 + (\beta^*)^2 \right) - i e^{\phi/4}  h_z \left( (\alpha^*)^2 - (\beta^*)^2 \right), \cr
0 =& - \frac{1}{4} F_0 e^{5\phi/4} \left( |\alpha|^2 - |\beta|^2 \right) + \frac{3}{4} \left( 2 c_1 - 3 c_2 + 3 c_3 \right)
\cr& - \frac{e^{\phi/4}}{8} \left[ g_z \left( (\alpha^*)^2 + (\beta^*)^2 \right) + i h_z \left( (\alpha^*)^2 - (\beta^*)^2 \right) \right]
 - \frac{e^{-\phi/2}}{2} b_z \alpha^* \beta^*.
\end{align}
There is an equation involving the derivative of the dilaton
\begin{align}
\label{eq:dil}
0&= \frac{1}{\rho} \p_z \phi \alpha^*+\frac{5}{4}F_0e^{5\phi/4}\beta + \frac{i}{4}e^{\phi/4}h_z\beta^* -\frac{1}{4}e^{\phi/4} g_z \beta^* + \frac{3}{4}F_0e^{3\phi/4}\frac{b_0}{f_1^2}\beta - \frac{1}{2} e^{-\phi/2} b_z \alpha^*\,.
\end{align}
Finally, there are four equations which are differential in $\alpha$ and $\beta$
\begin{align}
\label{eq:diff}
0&=\frac{1}{\rho} \p_z \alpha-\frac{1}{2} \frac{\p_z \rho}{\rho^2} \ \alpha +\frac{i}{4}e^{\phi/4}h_z\beta +\frac{1}{4}e^{\phi/4} g_z \beta - \frac{1}{4} e^{-\phi/2} b_z \alpha, \nonumber\\
0&=\frac{1}{\rho} \p_z \beta-\frac{1}{2} \frac{\p_z \rho}{\rho^2} \beta +\frac{i}{4}e^{\phi/4}h_z\alpha -\frac{1}{4}e^{\phi/4} g_z \alpha + \frac{1}{4} e^{-\phi/2} b_z \beta, \nonumber\\
0&=\frac{1}{\rho} \p_{z} \alpha^*+\frac{1}{2} \frac{\p_z \rho}{\rho^2} \alpha^* -\frac{1}{16}F_0e^{5\phi/4}\beta-\frac{i}{16}e^{\phi/4}h_{z}\beta^* +\frac{1}{16}e^{\phi/4}g_{z}\beta^* \cr &\qquad +\frac{1}{16}F_0e^{3\phi/4}\frac{b_0}{f_1^2}\beta - \frac{1}{8} e^{-\phi/2} b_z \alpha^*,\nonumber\\
0&=\frac{1}{\rho} \p_{z} \beta^*+\frac{1}{2} \frac{\p_z \rho}{\rho^2} \beta^* -\frac{1}{16}F_0e^{5\phi/4}\alpha -\frac{i}{16}e^{\phi/4}h_{z}\alpha^* -\frac{1}{16}e^{\phi/4}g_{z}\alpha^* \cr&\qquad -\frac{1}{16}F_0e^{3\phi/4}\frac{b_0}{f_1^2}\alpha + \frac{1}{8} e^{-\phi/2} b_z \beta^*.
\end{align}
The quantities $b_0$ and $b_z$ appear only algebraically and can be eliminated.  The remaining variables are then given by $\phi$, $\varphi_1$, $\varphi_2$, $\rho$ and the two spinor components $\alpha$ and $\beta$.  In total this gives $4$ real variables and $2$ complex variables.  The equations possess a conformal symmetry, with the following weighting
\begin{align}
&\alpha,\beta: (\frac{1}{4}, -\frac{1}{4})&
&\rho: (\frac{1}{2},\frac{1}{2})&
&g_z, h_z, b_z: (1,0)&
&b_0, \phi, \varphi_1, \varphi_2: (0,0)\,.&
\end{align}
We also note that the equations have a real scaling symmetry under $\alpha \rightarrow \lambda \alpha$, $\beta \rightarrow \lambda \beta$ and $c_i \rightarrow \lambda^2 c_i$, where $\lambda$ is an arbitrary real number.  This allows us to fix one of the $c_i$ without loss of generality, by absorbing it into the definition of $\alpha$ and $\beta$.  There is also a symmetry under $z \rightarrow - z$, $c_i \rightarrow -c_i$ and $F_0 \rightarrow - F_0$.

Next we consider the case $F_0 = 0$.  As discussed below \eqref{BPSeqns}, in order to make the limit $F_0 = 0$ well defined, we first make the replacement $B_{2} \rightarrow B_{2} - F_0^{-1} F_{2}$.  This introduces the closed two form $F_{2} = b_1 \hat e^{01}$, where $b_1$ is a constant.  Note that $A_{1} \wedge F_{2} = 0$ and so $C_{3}$ remains unmodified.  Sending $B_2 \rightarrow B_2 - F_2/F_0$ amounts to sending $b_0 \rightarrow b_0 - b_1/F_0$.  The massless limit is then obtained by setting $F_0 = 0$.

\section{Enhanced supersymmetry: $\gamma = -1/2,-2$ and D4-branes}
\label{sec:four}

We consider the case of supersymmetry enhancement which occurs by setting $h_z=b_0=0$.  This corresponds to the second line in \eqref{eq:doublesusy} and we shall see that the supersymmetry is enhanced to $F(4;2) \times SO(3)$.  The first line is equivalent to this case.  The algebraic equations \eqref{algconsts} combine to give the constraint $5 c_1 - 7 c_2 + 8 c_3 = 0$, which together with $c_1 - 2 c_2 + 2 c_3 = 0$, implies
\begin{align}
c_1 = - c_2 = - \frac{2}{3} c_3\,.
\end{align}
We combine $\alpha$ times the first equation of \eqref{eq:diff} with $\beta$ times the second to obtain
\beq
\p_z \left(\frac{\alpha^2+\beta^2}{\rho}\right)=0\,,
\eeq
which implies the existence of a holomorphic (1,0)-form $\kappa$,
\beq\label{D4:DefKappa}
\bar{\kappa} = \frac{\rho}{\alpha^2+\beta^2}\,.
\eeq
The first and second algebraic constraints in \eqref{algconsts} are solved to give $\phi$ and $g_z$
\begin{align}
&e^{5\phi/4} = \frac{2(c_2+c_3)}{5F_{0}}\frac{1}{|\alpha|^2-|\beta|^2},&
&g_z = -2 (c_2+ c_3) e^{-\phi/4} \frac{\kappa}{\rho}\,.&
\end{align}

We now rewrite the four equations in \eqref{eq:diff} as follows.  First we differentiate the conjugate of \eqref{D4:DefKappa} and use the last two equations of \eqref{eq:diff} to eliminate derivatives of $\alpha^*$ and $\beta^*$.  This leads to a differential equation for $\rho$
\beq\label{D4:rhoeq}
\frac{1}{\kappa}\p_z \ln \frac{\rho^2}{|\kappa|^2} = \frac{c_2+c_3}{20}\frac{1}{|\alpha|^2-|\beta|^2} \frac{c_3 f_3}{\nu_3}\,.
\eeq
Using this, as well as the equations \eqref{eq:diff}, we find that the $\p_z \phi$ equation leads to the constraint $5 c_1 - 7 c_2 + 8 c_3 = 0$ and is automatically satisfied.  We rewrite the first two differential equations of \eqref{eq:diff} as
\bea\label{D4:diffeqs}
\frac{1}{\kappa}\p_z \left(\frac{\alpha\beta}{\rho}\right) &=& -\frac{1}{2}(c_2+c_3)\frac{\alpha^2-\beta^2}{\rho}\,,\nonumber\\
\frac{1}{\kappa}\p_z \left(\frac{\alpha^2-\beta^2}{\rho}\right) &=& 2(c_2+c_3)\frac{\alpha\beta}{\rho}\,.
\eea
For the final equation we use the combination
\beq\label{D4:alpha2-beta2}
\p_z (|\alpha|^2-|\beta|^2) =\frac{5}{8}(c_2+c_3)\frac{c_3 \kappa}{\nu_3} f_3\,.
\eeq

We now move on to solving the equations as follows.  The strategy will be to introduce $h$ and $\tilde{h}$ by $(c_2+c_3)\kappa=\p_z\tilde{h}=i \p_{z} h$, integrate the equations \eqref{D4:diffeqs} and use their results to find expression for $\alpha$ and $\beta$. This will introduce a new holomorphic function which we will fix in terms of $h$ and $\tilde{h}$ through equations \eqref{D4:rhoeq} and \eqref{D4:alpha2-beta2}. With $\alpha$ and $\beta$ fully determined we can then write down the full solution.

The general solution to \eqref{D4:diffeqs} is given by
\beq
\frac{\alpha\beta}{\rho} &=& -\frac{1}{2}\mathcal{F}_1\sin(\frac{\tilde{h}+ih}{2}) + \mathcal{F}_2\cos(\frac{\tilde{h}+ih}{2})\, ,\nonumber\\
\frac{\alpha^2-\beta^2}{\rho}&=& \mathcal{F}_1\cos(\frac{\tilde{h}+ih}{2}) +2\mathcal{F}_2\sin(\frac{\tilde{h}+ih}{2})\, ,
\eeq
where $\mathcal{F}_1$ and $\mathcal{F}_2$ are functions of $\bar{z}$ only. For later convenience, we redefine
\beq
\mathcal{F}_1=\frac{1}{2}\left(\frac{\bar{\omega}}{\bar{\kappa}}+\frac{1}{\bar{\omega}\bar{\kappa}}\right),\qquad \mathcal{F}_2= \frac{1}{4i}\left(\frac{\bar{\omega}}{\bar{\kappa}}-\frac{1}{\bar{\omega}\bar{\kappa}}\right)\,,
\eeq
where $\omega$ and $\kappa$ are arbitrary holomorphic functions.  We will show below that this definition is consistent with \eqref{D4:DefKappa}.  It is convenient to work with the complex combinations
\beq
\alpha+i\beta=\rho^{1/2}\left(\frac{\bar{\omega}}{\bar{\kappa}}\right)^{1/2} e^{\tfrac{h-i\tilde{h}}{4}},\qquad\alpha-i\beta=\rho^{1/2}\left(\frac{1}{\bar{\omega}\bar{\kappa}}\right)^{1/2} e^{-\tfrac{h-i\tilde{h}}{4}}\,.
\eeq
Multiplying these together reproduces \eqref{D4:DefKappa}.
We are now ready to turn to the differential equation for $\rho$, we will use this equation as well as $\p_z \bar{\omega}=0$ to constrain the holomorphic function $\omega$. Using the above expressions for $\alpha$ and $\beta$, \eqref{D4:rhoeq} becomes
\beq
\frac{1}{\kappa}\p_z \ln \frac{\rho^2}{|\kappa|^2}= \frac{i(c_2+c_3)}{20}\frac{\omega-\bar{\omega}e^{-i\tilde{h}}}{\omega+\bar{\omega}e^{-i\tilde{h}}}.
\eeq
Differentiating this equation with respect to $\bar{z}$, multiplying by $\kappa$ and requiring the result to be real leads to the condition
\beq
i \frac{1}{\bar{\kappa}}\p_{\bar{z}} |\omega|^2 + i \frac{1}{\kappa}\p_{z} |\omega|^2= 4(c_2+c_3)i\p_{\tilde{h}}|\omega|=0\,.
\eeq
The condition $\p_{\tilde{h}}|\omega|=0$ leads through $\p_z \bar{\omega}=0$ also to $\p_h\theta_{\omega}=0$, where we defined $\omega=|\omega|e^{i\theta_{\omega}}$.

Next we note that the differential equation for $|\alpha|^2-|\beta|^2$, \eqref{D4:alpha2-beta2}, can be combined with the $\rho$-equation, \eqref{D4:rhoeq}, to give
\beq
\p_z \ln\left(\frac{\rho^2}{|\kappa|^2}(|\alpha|^2-|\beta|^2)^{-2/25}\right)=0\,.
\eeq
This is easily integrated to give
\beq\label{D4:rhointermediatesol}
\frac{\rho^2}{|\kappa|^2}=A^{24/25}(|\alpha|-|\beta|^2)^{2/25} = A\cos^{1/12} \left(\theta_{\omega}+\frac{\tilde{h}}{2}\right)
\eeq
where $A$ is a real constant.  Turning to the differential equation for $\rho$, we require \eqref{D4:rhointermediatesol} to be a solution. Plugging in we find that $\theta_{\omega} = \tfrac{1}{10} \tilde{h}$ and also $|\omega|= e^{-\tfrac{1}{10} h}$.  Note that we have absorbed integration constants into the definitions of $h$ and $\tilde h$.

We have now solved the system and give the final expressions for the supergravity fields.  The metric factors are given by
\begin{align}
&f_1 = \frac{\nu_1}{c_1} A^{1/2} \cos^{1/24}(\tfrac{3}{5}\tilde{h}) \frac{1}{2}\left(e^{\tfrac{2}{5}h}+e^{-\tfrac{2}{5}h}\right),& &f_3=-\frac{\nu_3}{c_3}A^{1/2}\cos^{1/24}(\tfrac{3}{5}\tilde{h})\sin(\tfrac{3}{5}\tilde{h}),&\cr
\label{D4:warpfactSol1}
&f_2 = -\frac{\nu_2}{c_2} A^{1/2} \cos^{1/24}(\tfrac{3}{5}\tilde{h}) \frac{1}{2}\left(e^{\tfrac{2}{5}h}-e^{-\tfrac{2}{5}h}\right),&
&\frac{\rho^2}{|\kappa|^2}=\cos^{1/24}(\tfrac{3}{5}\tilde{h}).&
\end{align}
The dilaton and flux are given by
\begin{align}
&e^{5\phi/4}=\frac{2}{5}\frac{c_2+c_3}{F_{0}}A^{-1/2}\cos^{-25/24}(\tfrac{3}{5}\tilde{h}),&
&\varphi_2 = \frac{5}{2}\frac{A^{3/2}}{c_3^3}\left(\frac{2}{5}\frac{c_2+c_3}{F_{0}A^{1/2}}\right)^{-1/5}\cos^{4/3}(\tfrac{3}{5}\tilde{h})\,.&
\end{align}
with $g_z = \nu_3 (\p_z \varphi_2)/f_3^3 \rho$.

This metric reproduces exactly the $AdS_6$ solution of \cite{Brandhuber:1999np}.  This can be seen by introducing the new coordinates $x$ and $y$ by
$h=5x/2$ and $\tilde{h}=5y/3$.  We also introduce the overall radius $R$ by $A^{1/2}=\frac{2}{5}(c_2+c_3)R$. In these coordinates, the metric is given by
\beq
\label{ads6sola}
ds^2 = R^2\cos^{\frac{1}{12}}y \left[\left(dx^2 +\cosh^2 x \, ds^2_{AdS_2} + \sinh^2 x \, ds^2_{S^3}\right) + \frac{4}{9}\left(dy^2+\sin^2y\, ds^2_{S^3}\right)\right].
\eeq
The terms in the first set of parenthesis combine into an $AdS_2\times S^3$ slicing of $AdS_6$, whereas the terms in the second set combine to make an $S^4$. The dilaton and flux in these coordinates are
\beq
\label{ads6solb}
e^{\phi/4}= (RF_{0})^{-1/5} \cos^{-5/24} y,\qquad \varphi_2= \frac{4}{9}R^3 (RF_{0})^{1/5} \cos^{4/3} y.
\eeq

\subsection{The case $F_{0}=0$}
There are no solutions with $F_{0}=0$.  To see this we first take the $F_0 \rightarrow 0$ limit of the BPS equations of section \ref{sec:summBPS}.  Since we are interested in solutions with the only non-vanishing flux given by $g_z$, we may simply take $F_0 = 0$.
In this case, the algebraic equations \eqref{algconsts} become
\bea
0&=&2c_1-3c_2+3c_3,\\
0&=&2(c_2+c_3) + e^{\phi/4}\big((\alpha^*)^2+(\beta^*)^2\big)\nonumber\\
0&=& 6(2c_1-3c_2+3c_3) - e^{\phi/4}\big((\alpha^*)^2+(\beta^*)^2\big).\nonumber
\eea
Along with the original constraints, we obtain a total of three constraints on the $c_i$
\beq
c_2+c_3=0,\qquad 2c_1-3c_2+3c_3=0,\qquad c_1-2c_2+2c_3=0,
\eeq
from which it follows that $c_i=0$.

\section{Enhanced supersymmetry: $\gamma = 1$ and fundamental strings}
\label{sec:five}

In this section we consider the case in which supersymmetry is enhanced by setting $g_z = h_z = 0$.  This corresponds to the third line in \eqref{eq:doublesusy} and we shall see that the supersymmetry is enhanced to $OSp(8|2,\mathbb{R})$.

The second equation of \eqref{algconsts} yields the condition $c_2 = -c_3$, which together with the constraint $c_1 - 2 c_2 + 2 c_3 = 0$, implies
$c_1 = 4 c_2 = -4 c_3$.  Next, we combine $\beta$ times the first equation of \eqref{eq:diff} with $\alpha$ times the second to yield the vanishing of a total derivative, $\partial_z \ln\left(\alpha \beta/\rho\right) = 0$.  This implies the existence of a holomorphic $(1,0)$-form $\kappa$,
\begin{align}
\bar{\kappa}=c_2 \frac{\rho}{\alpha \beta}\, .
\end{align}
The factor of $c_2$ has been chosen for convenience.
We write the remaining three independent equations of \eqref{eq:diff} as
\begin{align}
\label{eq:reddiff}
\p_z b_0 =&2f_1^2 e^{\phi/2}\partial_z \ln \left(\frac{\alpha}{\beta}\right)\,, \cr
\partial_z\ln\left(\alpha(\beta^*)^2\sqrt{\rho}\right)=&
\frac{1}{8}\left(1+\frac{b_0}{f_1^2}e^{-\phi/2}\right)F_{0}e^{5\phi/4}\frac{\rho\alpha}{\beta^*}\,, \cr
\partial_z\ln\left(\beta(\alpha^*)^2\sqrt{\rho}\right)=&
\frac{1}{8}\left(1-\frac{b_0}{f_1^2}e^{-\phi/2}\right)F_{0}e^{5\phi/4}\frac{\rho\beta}{\alpha^*}\,,
\end{align}
where we have used $b_z = (\p_z b_0)/f_1^2 \rho$.  We solve these equations to give $b_0$ and $\phi$ as functions of the remaining variables.  The remaining equation for $\p_z b_0$ is then automatic.
After substituting in the above solutions for $b_0$ and $\phi$ and eliminating $\beta$ in terms of $\kappa$, the first equation of \eqref{algconsts} becomes
\begin{align}
\label{eq:alg15}
0 = |\kappa|^2 + 2 \bar \kappa \left[ \frac{|\kappa|^2 |\alpha|^4}{\rho^2} \p_z \ln \left(\frac{(\alpha^*)^2}{ \alpha} \rho^\frac{3}{2} \right)
- \frac{\rho^2}{|\kappa|^2 |\alpha|^4} \p_z \ln \left(\frac{\alpha}{\kappa^2 (\alpha^*)^2} \rho^\frac{5}{2}  \right) \right]\,.
\end{align}
Next we take the sum of the first equation in \eqref{algconsts} and third equation in \eqref{algconsts} and again eliminate $b_0$, $\phi$ and $\beta$ to obtain
\begin{align}
\label{eq:totder5}
0 = \partial_z \ln \left(\frac{\alpha^4}{\kappa^2(\alpha^*)^4}\right)-2 \kappa\,.
\end{align}
The final remaining equation is given by \eqref{eq:dil}.

We integrate \eqref{eq:totder5} as follows.  We parameterize the magnitude and phase of $\alpha$ by the quantities $A$ and $\theta$ as $\alpha=\sqrt{c_2 \rho}A^{1/4}e^{i\theta}/\sqrt{\bar \kappa}$.  We also introduce a real harmonic function $h$ by $i\partial_z h = \kappa$.  In terms of these quantities, \eqref{eq:totder5} becomes
\beq
h = 4\theta\, .
\eeq
With this we can write \eqref{eq:alg15} as a differential equation for $\rho$ in terms of $A$ and $\kappa$
\begin{align}\label{FSSRhoEq}
\p_z \ln \left(\frac{\rho^2}{|\kappa|^2}\right)  =& -\frac{1}{4} \left(\frac{3A^2-2A+3}{1-A^2}\right) \kappa
+ \frac{1}{4} \left(\frac{1+A^2}{1-A^2}\right) \p_z \ln A\, .
\end{align}
Using the above equation, we can eliminate $\rho$ from the equations and cast the remaining system as a pair of first order differential equations for $\phi$ and $A$.  The first is obtained the last two equations in \eqref{eq:reddiff} after eliminating $b_0$.  The second is the differential equation  \eqref{eq:dil}.  It will be slightly more convenient to introduce a new variable $G$ for the dilaton, defined by the equation
\begin{align}
e^\frac{5\phi}{2} = \frac{|\kappa|^2}{\rho^2} \frac{A G^2}{F_{0}^2}\, .
\end{align}
Note that $G$ must have the same sign as the product $\rho \, F_0$.  In terms of $G$ and $A$ the remaining first order system is given by
\begin{align}\label{FSSFinalDiffEq}
\p_z \ln \left(G \right) =&
-  \frac{5 A + 1}{1-A^2}  \p_z \ln(A)
- \frac{A^2 - 14 A + 1}{1-A^2} \kappa\, , \cr
\kappa G =&
- \frac{4}{1-A} \kappa
+ \frac{2}{1-A} \p_z \ln(A)\, .
\end{align}

Our approach will be to solve equations (\ref{FSSFinalDiffEq}) to obtain expressions for $A$ and $G$. We then integrate (\ref{FSSRhoEq}) to obtain $\rho$.  The metric factors, dilaton and fluxes are then determined uniquely in terms $G$, $A$, $\rho$ and $h$.  The metric factors are
\begin{align}
\label{fund:mets}
&f_1^2 = \frac{\rho^2}{16 |\kappa|^2} A \left(1 + A^{-1} \right)^2\, ,&
&f_2^2 = \frac{4 \rho^2}{|\kappa|^2} \sin^2 \left( \frac{h}{2}  \right)\,,&
&f_3^2 = \frac{4 \rho^2}{|\kappa|^2} \cos^2 \left( \frac{h}{2}  \right)\,,&
\end{align}
while the dilaton $\phi$ and flux $b_0$ are given by
\begin{align}
\label{fund:dils}
&e^\frac{5\phi}{2} = \frac{|\kappa|^2}{\rho^2} \frac{A G^2}{F_{0}^2}\,,&
&b_0 = \frac{e^{-\frac{3\phi}{4}}}{4 F_{0}}\frac{\rho}{|\kappa|}\left(A^{-1/2}+A^{1/2}\right)\left(1-\frac{1-A}{4}G\right)\,.&
\end{align}

From these expressions, we can see the presence of an $SO(8)$ symmetry as follows.  By a conformal transformation, we may pick $h$ as a coordinate.  Introducing the dual coordinate $\tilde h$ as $\kappa = \p_z \tilde h$, this corresponds to the choice $z = \tilde h + i h$ with $\kappa = 1/2$.  With this choice of coordinates, the differential equations \eqref{FSSFinalDiffEq} imply that $A$ and $G$ depend only on $\tilde h$.  Similarly, \eqref{FSSRhoEq} implies $\rho$ only depends on $\tilde h$.  As a result, we find the metric is given by
\begin{align}
ds^2 = f_1^2 ds^2_{AdS_2} + 4 \rho^2 d \tilde h^2 + 16 \rho^2 \left[\frac{dh^2}{4} + \sin^2\left(\frac{h}{2}\right) ds^2_{\mathbb{S}^3} + \cos^2\left(\frac{h}{2}\right) ds^2_{\mathbb{S}^3}\right]\,.
\end{align}
The metric in brackets is that of $7$-sphere with unit radius, whose isometry group is $SO(8)$.  As a consequence, we find that the full symmetry group is $OSp(8|2,\mathbb{R})$, as advertised.

\subsection{Linear dilaton}
\label{sec:simpsol}
We present a simple solution to the equations \eqref{FSSFinalDiffEq} obtained by taking $A$ constant.  Assuming $A$ is constant leads to the condition $A^2 - 14 A + 1 = 0$.  This has two solutions $A = 7 \pm 4 \sqrt{3}$ with the corresponding $G$ given by $G = -2 \pm (4/\sqrt{3})$.
For constant $A$, the $\rho$ equation \eqref{FSSRhoEq} becomes
\begin{align}
\p_z \ln \left( \frac{\rho^2}{|\kappa|^2} \right) = \pm \frac{5}{4 \sqrt{3}} \kappa.
\end{align}
To integrate, we introduce $\tilde h$ so that $\kappa = \p_z \tilde h$.  Integrating then gives
$\rho^2 = L^2 |\kappa|^2 e^{\pm \frac{5}{4 \sqrt{3}} \tilde h}$, where $L$ is an integration constant.

We are still free to make conformal transformations and by a local change of coordinates, we can choose $z = \tilde h + i h$.  This corresponds to using a conformal transformation to set $\kappa = 1/2$.  Finally, it will be convenient to introduce the rescaled variables $x$ and $\theta$ for $\tilde h$ and $h$ so that $z  = \pm (4\sqrt{3}/5) x + 2 i \theta$.

The dilaton and flux are given by
\begin{align}
&e^{-5 \phi/2} =  \frac{3}{4} L^2 F_{0}^2 e^{x}\,,&
&b_0 = 3^{3/10}\left(\frac{2 L}{F_{0}}\right)^{2/5}e^{\frac{4}{5}x}\,.&
\end{align}
The metric factors are given by
\begin{align}
&f_1^2 = L^2 e^{x},&
&f_2^2 = 4 L^2 \sin^2(\theta) e^{x},&
&f_3^2 = 4 L^2 \cos^2(\theta) e^{x},&
\end{align}
and the metric becomes
\begin{align}
ds^2= L^2 e^{x}\left(\frac{48}{25} dx^2+ ds^2_{AdS_2} + 4\,ds^2_{S^7}\right)\,.
\end{align}

As $x \rightarrow \infty$, the geometry becomes asymptotically flat and the string coupling tends to zero.  As $x \rightarrow -\infty$, the geometry becomes strongly curved and the string coupling becomes large.  Note that in $(\tilde h, h)$-coordinates, the two solutions are simply mirrors of each other.

\subsection{General solutions}
\label{ssgensols}
In this section, we study general solutions to the system of equations \eqref{FSSFinalDiffEq}.  The differential equations have singularities at $A = 0,1,\infty$.  We start by analyzing the solution in the neighborhood of each of these points.

We start by obtaining a solution to \eqref{FSSFinalDiffEq} in the large $A$ limit.  Using the second equation of (\ref{FSSFinalDiffEq}) to eliminate $\p_z \ln A$ in the first equation and then dropping terms which are sub-leading in the large $A$ limit, the equations reduce to
\begin{align}
\label{eq:diffexpanded}
&\p_z \ln (A) \sim - \frac{A}{2} \kappa G + 2 \kappa\,,&
&\p_z \ln G \sim - \frac{5}{2} \kappa G + \kappa\,.&
\end{align}
The right equation can be easily integrated by again introducing $\tilde h$ as $\p_z \tilde h = \kappa$, to give
\begin{align}
G \sim \frac{2 \, \CI e^{\tilde h}}{1+5 \, \CI e^{\tilde h}},
\end{align}
where $\CI$ is a real integration constant.

Turning now to the first equation, we first assume that $G$ is finite in the large $A$ limit.  In this case, we can neglect the second term on the left and side and the equation is easily integrated to give
\begin{align}
\frac{1}{A} \sim \frac{1}{5}\ln \left(1+5 \CI e^{\tilde h}\right)+ \CII,
\end{align}
where $\CII$ is another real integration constant.  We introduce a new coordinate $\lambda$ by $\tilde{h}=\ln [\left(e^{5\lambda-5\CII}-1\right)/5 \CI ]$.  Expanding around $\lambda = 0$ leads to the following asymptotic behavior
\begin{align}
\label{leadingsol}
&{\rm Case \, I}:&
&A \sim \lambda^{-1}\,,&
&G \sim 6 \lambda\,,& &(\CII = 0)& \cr
&{\rm Case \, II}:&
&A \sim \lambda^{-1}\,,&
&G \sim G_0\,,& &(\CII \neq 0)&
\end{align}
where $G_0 = 2(1-e^{5 \CII})/5$.  For the case $\CII = 0$, we find that $A G$ is of order one and so we must keep the second term in the first equation of \eqref{eq:diffexpanded}, which then yields the correct factor of $6$.

By a local change of coordinates, we can again pick $z = \tilde h + i h$, corresponding to setting $\kappa = 1/2$.  Expanding the differential equation \eqref{FSSRhoEq} for $\rho$ to leading order and integrating, we find that the metric on $\Sigma_2$ has the asymptotic form
\begin{align}
&{\rm Case \, I}:&
&4 \rho^2 (d\tilde h^2 + dh^2) \sim L^2 \frac{d \lambda^2}{\lambda} + L^2 \lambda dh^2\,,& \cr
&{\rm Case \, II}:&
&4 \rho^2 (d\tilde h^2 + dh^2) \sim L^2 \lambda^\frac{1}{4} \left( dh^2 + \frac{4}{G_0^2} d\lambda^2 \right)\,,&
\end{align}
where $L$ is an integration constant.  The warp factors behave as
\begin{align}
&{\rm Case \, I}:&
&f_1^2 \sim \frac{L^2}{16}\,,&
&f_2^2 \sim 4L^2 \lambda \sin^2 \left( \frac{h}{2}  \right)\,,&
&f_3^2 \sim 4L^2 \lambda \cos^2 \left( \frac{h}{2}  \right)\,,& \cr
&{\rm Case \, II}:&
&f_1^2 \sim \frac{L^2}{16 \lambda^\frac{3}{4}} \,,&
&f_2^2 \sim 4L^2 \lambda^\frac{1}{4} \sin^2 \left( \frac{h}{2}  \right)\,,&
&f_3^2 \sim 4L^2 \lambda^\frac{1}{4} \cos^2 \left( \frac{h}{2}  \right)\,.&
\end{align}
For case I, we change coordinates to $h = 2 \theta$ and $\lambda = r^2$ and for case II, we introduce $h = 2 \theta$.  In these coordinates, the asymptotic metrics take the form
\begin{align}
\label{asymc1c2a}
&{\rm Case \, I}:&
&ds^2 \sim L^2 \left(\frac{1}{16} ds^2_{AdS_2} + 4 dr^2 + 4 r^2 ds^2_{\mathbb{S}^7} \right)\,,&
\cr
&{\rm Case \, II}:&
&ds^2 \sim L^2 \lambda^\frac{1}{4} \left(\frac{1}{16\lambda} ds^2_{AdS_2} + \frac{4}{G_0^2} d\lambda^2 + 4 ds^2_{\mathbb{S}^7} \right)\,.&
\end{align}
The asymptotic values of the dilaton and flux are given by
\begin{align}
\label{asymc1c2b}
&{\rm Case \, I}:&
&e^{\frac{5 \phi}{2}} \sim \frac{36}{L^2 F_0^2}\,,&
&b_0 \sim \frac{5}{8} \left(\frac{L^8}{6^3 F_0^2}\right)^\frac{1}{5}\,,& \cr
&{\rm Case \, II}:&
&e^{\frac{5 \phi}{2}} \sim \frac{G_0^2}{L^2 F_0^2 \lambda^\frac{5}{4}} \,,&
&b_0 \sim \frac{1}{16 \lambda} \left(\frac{G_0^2 L^8}{F_0^2}\right)^\frac{1}{5}\,.&
\end{align}

For case I, we see that as $r \rightarrow 0$ the geometry caps off smoothly.  Furthermore, the dilaton, $\phi$, and flux, $b_0$, both remain finite.  For case II, the geometry is singular as $\lambda \rightarrow 0$, while the dilaton, $\phi$, and flux, $b_0$, both diverge.  We note that the metric is regular in string frame, although the coupling still diverges.\footnote{This is easily seen by recalling that $ds^2_{\rm string} = e^{\phi/2} ds^2_{\rm Einstein}$.}

Next we examine the solution near $A = 0$.  Assuming $A \sim 0$ and eliminating $G$ in favor of a second order equation for $A$, we obtain the approximate equation $2 A + \p_h A - \p_h^2 A = 0$, whose general solution is given by $A \sim \CIII e^{-\tilde h} + \CIV e^{2\tilde h}$.  Since we are in the $A \sim 0$ approximation, this equation implies the following allowed behaviors: $A \sim e^{2\tilde h}$ as $\tilde h \rightarrow -\infty$, $A \sim e^{-\tilde h}$ as $\tilde h \rightarrow \infty$ or $A$ has a first order zero in $\tilde h$.  The first case is an exact solution with $G=0$.  This leads to a solution with $e^{\phi} = 0$ and a divergent $b_0$.  The other two cases have the following asymptotics
\begin{align}
\label{leadingsol2}
&{\rm Case \, III}:&
&A \sim \lambda \,,&
&G \sim -6 \,,&
&\left(A \sim e^{-\tilde h}\right)&\cr
&{\rm Case \, IV}:&
&A \sim \lambda\,,&
&G \sim G_0 \lambda^{-1}\,,&
&\left(A \sim \frac{G_0}{2} \tilde h - \frac{G_0}{2} \tilde h_0 \right)&
\end{align}
where for case III, we introduced $\lambda$ by $\tilde h = - \ln \lambda$ and for case IV, $G_0$ and $\tilde h_0$ are integration constants and we have introduced $\lambda$ by $\lambda = G_0 (\tilde h - \tilde h_0)/2$.  It turns out that the asymptotic geometry takes the same form as in the $A \sim \infty$ cases.  Namely case III leads to the same asymptotics given in \eqref{asymc1c2a} and \eqref{asymc1c2b} for case I, while case IV leads to the same asymptotics for case II.

For the special point $A=1$, we find that $A$ admits a series expansion as a polynomial in $h$.  $G$ can have one of either two behaviors.  Either it admits a regular series expansion, with the value of $G$ arbitrary at $A=1$ or $G$ has a linear divergence at $A=1$ such that
\begin{align}
\label{leadingsol3}
&{\rm Case \, V}:&
&A \sim 1 \,,&
&G \sim \frac{2}{3} \lambda^{-1} \,,&
\end{align}
where $\lambda = (\tilde h - \tilde h_0)$ and $\tilde h_0$ is the location of $A=1$.  For this case, we change coordinates to $h = 2 \theta$ and $\lambda = 2 r$.  In these coordinates, the asymptotic metrics take the form
\begin{align}
\label{asymVa}
&{\rm Case \, V}:&
&ds^2 \sim L^2 r^\frac{1}{12} \left( ds^2_{AdS_2} + 16 dr^2 + 16 ds^2_{\mathbb{S}^7} \right)\,.&
\end{align}
The asymptotic values of the dilaton and flux are given by
\begin{align}
\label{asymVb}
&{\rm Case \, V}:&
&e^{\frac{5 \phi}{2}} \sim \frac{1}{26 L^2 F_0^2 r^\frac{25}{12}}\,,&
&b_0 \sim \frac{5}{4} \left(\frac{6^3 L^8}{ F_0^2}\right)^\frac{1}{5} r^\frac{2}{3}\,.&
\end{align}
This solution is singular as $r \rightarrow 0$.  However, we observe that the singularity is of the same type as that which occurs for the $AdS_6$ solution given in \eqref{ads6sola} and \eqref{ads6solb}.  In the $AdS_6$ case, the singularity was attributed to the presence of an O8-plane \cite{Brandhuber:1999np}.  Since we observe the same singularity structure, we interpret this solution as describing a fundamental string ending on an O8-plane.  The first type of behavior corresponds to a regular interior point and yields the behavior given in figure \ref{case2m}.

The above analysis parallels nicely with the brane picture.  We associate the case I behavior with a string ending on a stack of D8-branes from the left, where we have used $\kappa$ to define the orientation.  The case III behavior can be associated with a fundamental string ending on a stack of D8-branes from the right.  The case V behavior contains two disconnected solutions.  For the first, we take $h > h_0$ with $G>0$, this corresponds to a string ending on an O8-plane from the left and for the second we take $h < h_0$ with $G < 0$, corresponding to a string ending on an O8-plane from the right.  Finally, we conjecture that the simple solution of section \ref{sec:simpsol} can be associated with an infinite string in the presence of a non-zero Roman's mass.

\subsection{Numerics}

For each of the cases in the previous section, one can work out the series solution to any finite order.  However, we find the series expansion always breaks down for some finite value of $\tilde h$.  In order to understand the global structure of the solutions, we therefore solve the differential equations numerically.  To do so, we first find approximate series solutions for each of the singular points, corresponding to cases I-V of the previous section.  We use these series solutions to generate initial data for $A$ and $G$ away from the singular points.  Finally, we use this initial data to numerically solve the pair of differential equations given in \eqref{FSSFinalDiffEq}.

\begin{figure}[ht]
\begin{subfigure}[b]{0.5\textwidth}
\centering
\includegraphics[width=0.9\linewidth]{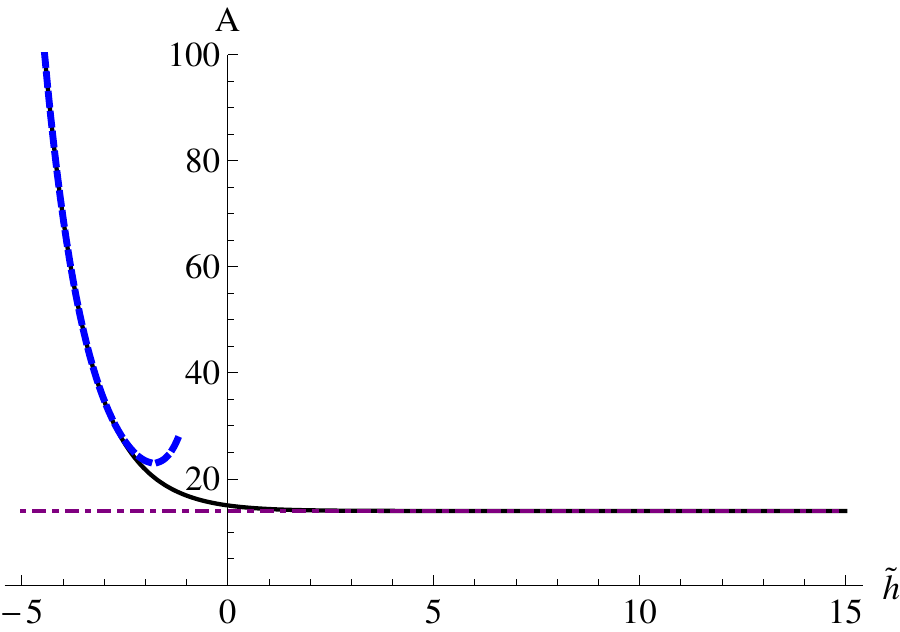}
\end{subfigure}
\begin{subfigure}[b]{0.5\textwidth}
\centering
\includegraphics[width=0.9\linewidth]{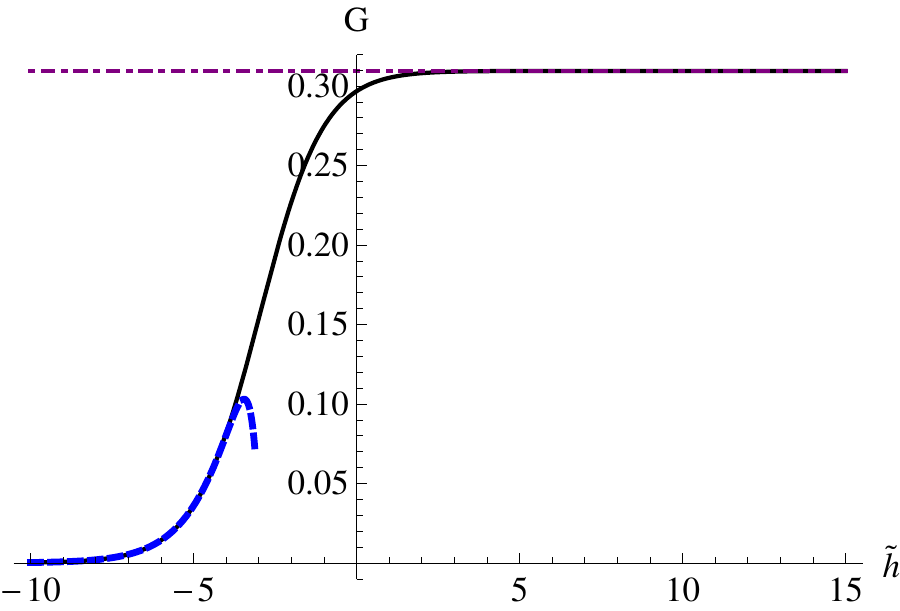}
\end{subfigure}
\caption{Plot of $A$ (left) and $G$ (right) for case I.  The solid line is the numerically generated solution. The dot-dashed horizontal line is the constant solution of section \ref{sec:simpsol}.  The dashed line is the series solution \eqref{CIIzerisersol} truncated at order $n=2$.}
\label{CIIzero}
\end{figure}

For case I, the differential equations admit the series solution
\begin{align}
\label{CIIzerisersol}
&A=a_{-1}e^{-\tilde{h}} + a_0 + a_1 e^{\tilde{h}} + \ldots = \sum_{n=-1}^{\infty} a_n e^{n\tilde{h}}\,,& \cr
&G= g_1 e^{\tilde{h}} + g_2 e^{2\tilde{h}}+\ldots = \sum_{n=1}^{\infty} g_n e^{n\tilde{h}}\,,&
\end{align}
where the coefficients $a_i$ and $g_i$ are constants and by a choice of coordinates we can set $a_{-1} = 1$.  The equations (\ref{FSSFinalDiffEq}) can be written as
\begin{align}
\label{eq:diffeqredsimp}
0 =& A(1-A^2)\partial_{\tilde{h}}G + G (5A+1) \partial_{\tilde{h}} A + A (A^2-14A+1)G\,,\cr
0=& A(1-A)G + 4A - 2 \partial_{\tilde{h}} A\,,
\end{align}
where we have used $\p_z \tilde h = \kappa$ and dropped an overall factor of $\kappa$ from the equations.  Plugging in the series expansion, one can recursively solve for all the remaining coefficients in terms of $a_{-1}$.  Since the differential equations are non-linear, it is difficult to obtain a closed form expression for the recurrence relation.  However, one can explicitly solve for the coefficients to any given order.  We have used MATHEMATICA to solve for the first fifty coefficients, with the first few given by
\begin{align}
&a_{-1}=1,& &a_{0}=15,& &a_{1}=-12,& &a_{2} = 144,&\cr
&g_{1}=6,& &g_{2}=-114& &g_3=2166& &g_4=-41250.&
\end{align}
Near $\tilde{h} \sim -3$, the series expansion exhibits rapid oscillations and appears to break down.  To obtain the behavior beyond this point, we solve the equations \eqref{eq:diffeqredsimp} numerically.  We use the series expansion to generate the initial data.  Starting at $\tilde{h}_0 = -15$, we find from the series solution that $A(\tilde{h}_0)=3.27\times 10^6$ and $G(\tilde{h}_0)=1.84\times 10^{-6}$.  The result of this numerical solution is shown in figure \ref{CIIzero}.  We see a nice agreement between the numerical solution and the series solution up to $\tilde h \sim -3$.  Beyond that we see the numerical solution smoothly interpolates to the constant $A$ and $G$ solution of section \ref{sec:simpsol}.  We conclude that case I, provides a smooth, weakly curved geometry.

\begin{figure}[ht]
\begin{subfigure}[b]{0.5\textwidth}
\centering
\includegraphics[width=.9\linewidth]{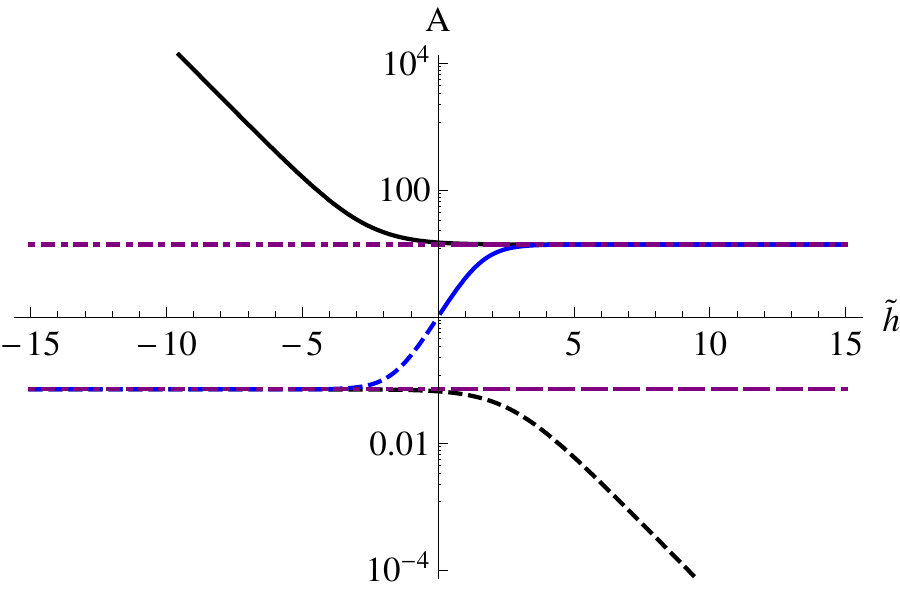}
\end{subfigure}
\begin{subfigure}[b]{0.5\textwidth}
\centering
\includegraphics[width=.9\textwidth]{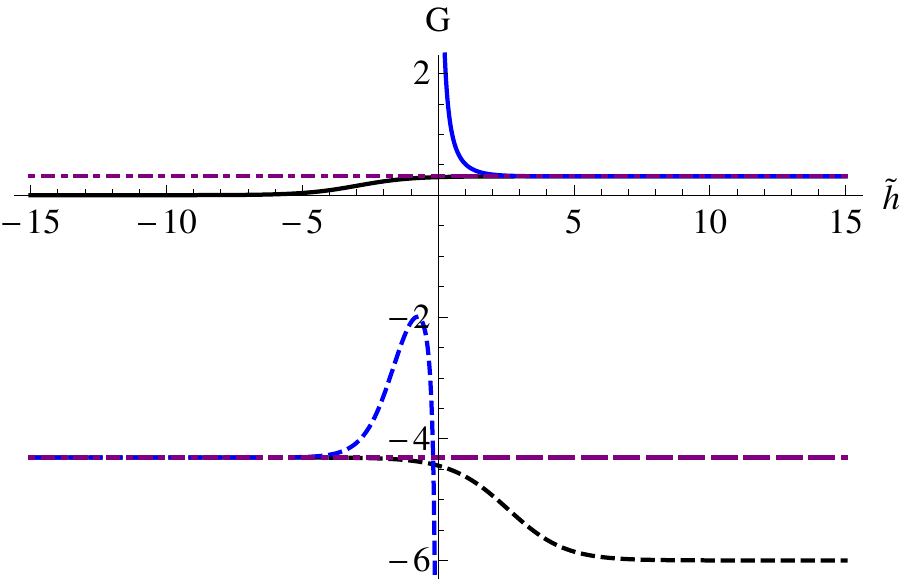}
\end{subfigure}
\caption{Plot of $A$ (left) and $G$ (right) versus $\tilde h$.  The black solid (dashed) line corresponds to case I (III) while the blue solid (dashed) line corresponds to the right (left) extension of case V. The dot-dashed horizontal line is the constant solution of section \ref{sec:simpsol}.}
\label{case135}
\end{figure}

The remaining cases proceed in a similar manner.  For case III, there is a unique numerical solution as in case I.  For case V, two numerical solutions can be generated, one from taking data to the right of the $A=1$ point and the other by taking data from the left.  The results of cases I, III and V are shown together in figure \ref{case135}.  In all three cases, the solutions asymptote to the constant $A$ and $G$ solution of section \ref{sec:simpsol}.  Note that case III is a reflected version of case I, while the left and right solutions of case V are reflections of each other.

\begin{figure}[ht]
\begin{subfigure}[b]{0.5\textwidth}
\centering
\includegraphics[width=.95\linewidth]{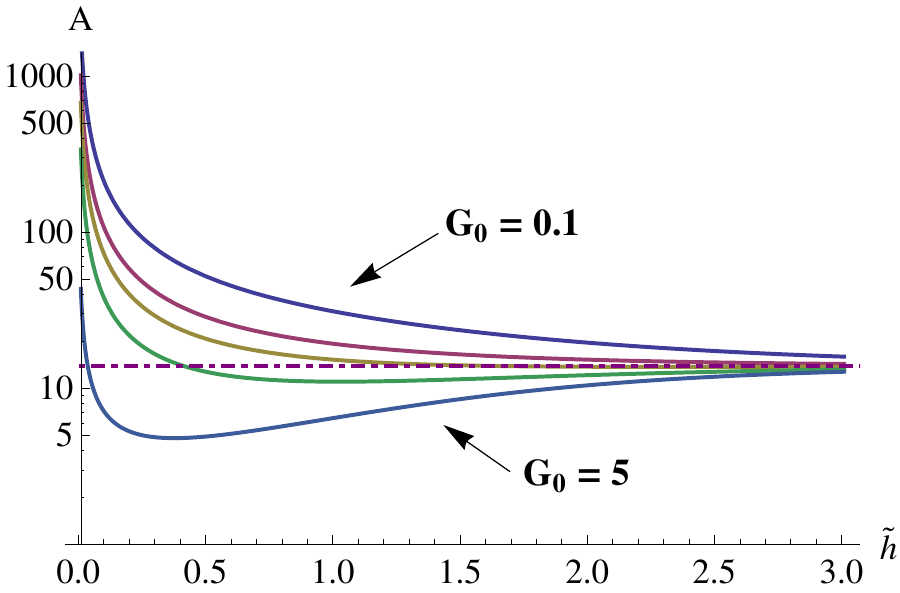}
\end{subfigure}
\begin{subfigure}[b]{0.5\textwidth}
\centering
\includegraphics[width=.95\linewidth]{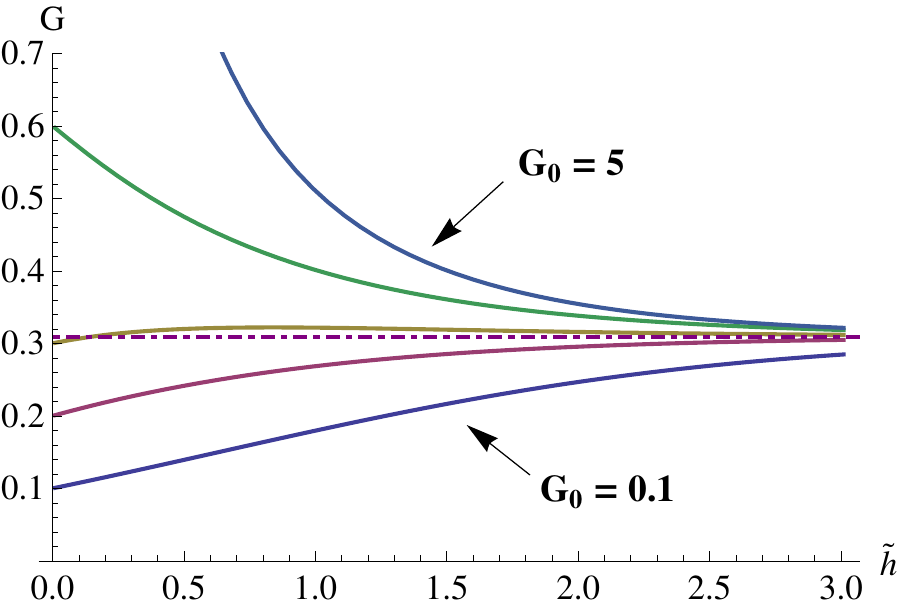}
\end{subfigure}
\caption{Plot of $A$ (left) and $G$ (right) versus $\tilde h$ for case II $G_0 = (0.1,0.2,0.3,0.6,5)$.  The dot-dashed horizontal line is the constant $G$ solution of section \ref{sec:simpsol}.}
\label{case2}
\end{figure}

\begin{figure}[ht]
\begin{subfigure}[b]{0.5\textwidth}
\centering
\includegraphics[width=.95\linewidth]{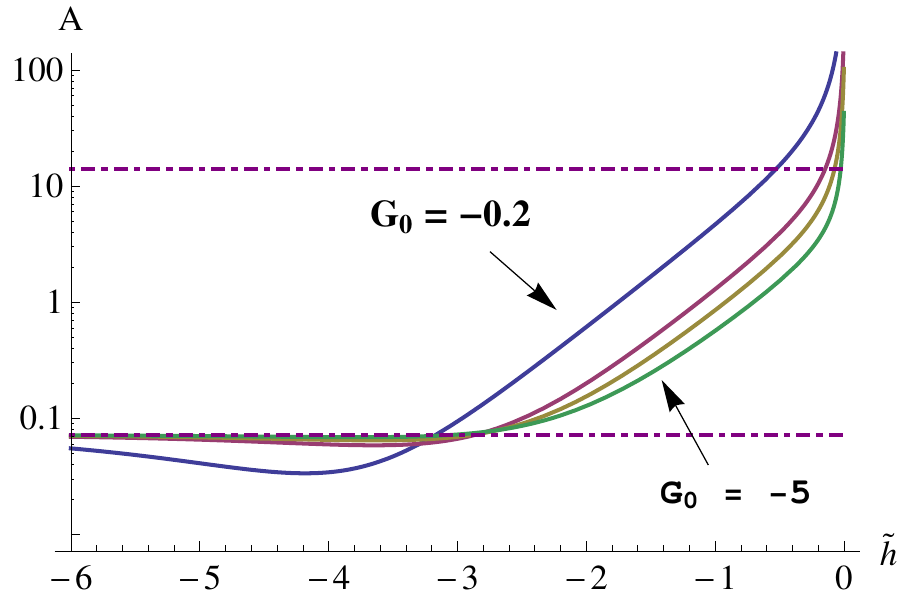}
\end{subfigure}
\begin{subfigure}[b]{0.5\textwidth}
\centering
\includegraphics[width=.95\linewidth]{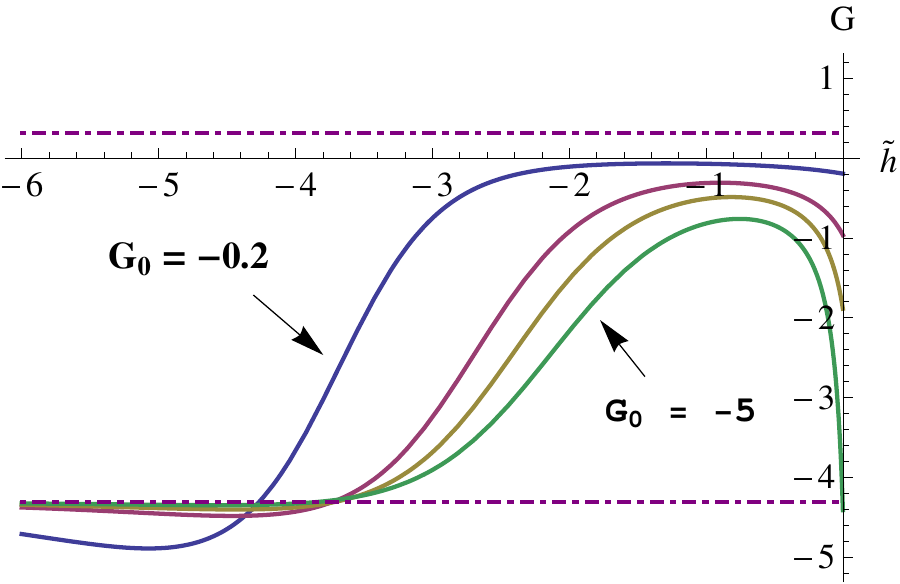}
\end{subfigure}
\caption{Plot of $A$ (left) and $G$ (right) versus $\tilde h$ for case II with $G_0 = (-0.2,-1,-2,-5)$.  The dot-dashed horizontal lines are the constant $G$ solution of section \ref{sec:simpsol}.}
\label{case2m}
\end{figure}

For case II, there is a one parameter family of solutions, which are shown in figure \ref{case2} for positive values of $G_0$ and in figure \ref{case2m} for negative values of $G_0$.  Again, we observe that the solution asymptotes to the constant $A$ and $G$ solution.  Case IV is a reflected version of case II.  We note that cases II and IV are both singular and do not necessarily have a brane interpretation.

In all cases, we observe that the solutions asymptote to the constant $A$ and $G$ solution.  This is consistent with the brane interpretation of the previous section.  Namely, the fact that all the solutions asymptote to the constant $A$ and $G$ solution corresponds to the fact that the fundamental strings are all semi-infinite.

\subsection{Patching solutions with jumping $F_0$}

In this section, we consider D8-brane domain walls, where the value of $F_0$ jumps across the D8-brane.  The physical fields are required to be continuous across the D8-brane \cite{Apruzzi:2013yva}.  In particular, the metric and gauge potentials are required to be continuous functions.  We parameterize the jump with a real parameter $\lambda$ such that $F_0^- = \lambda F_0^+$.  Where $F_0^{-(+)}$ is the value of $F_0$ on the left (right) side of the D8-branes.  Assuming that $\rho$ is continuous, the expressions \eqref{fund:mets} for the $f_i$ are continuous if and only if $A$ is continuous.  For $\rho$, we note that it is determined by the differential equation \eqref{FSSRhoEq} and we may always choose the integration constant so that $\rho$ is continuous.  The expression for the dilaton in \eqref{fund:dils} is continuous if and only if we assume $G$ jumps across the domain wall so that $G^- = \lambda G^+$.

At first sight, it seems that $B$ cannot be made continuous across the interface.  However, this is due to our choice of gauge for the $B$ field.  We first introduce a constant parameter $b_1$ and make the gauge transformation given by sending
$B_2 \rightarrow B_2 - (b_1/F_0) \hat e^{01}$ and $F_2 \rightarrow b_1 \hat e^{01}$.  This amounts to introducing $F_2$ and taking
\begin{align}
b_0 = \frac{e^{-\frac{3\phi}{4}}}{4 F_{0}}\frac{\rho}{|\kappa|}\left(A^{-1/2}+A^{1/2}\right)\left(1-\frac{1-A}{4}G\right) - \frac{b_1}{F_0}
\end{align}
Note that $C_3$ does not transform since $A_1 \wedge F_2 = 0$.  Working in the $h$ and $\tilde h$ coordinates, we consider inserting the D8-brane at some value of $\tilde h$, which we denote by $\tilde h_1$.  In order to obtain a solution with a continuous $B_2$ and $F_2$ we take
\begin{align}
b_1 = \frac{e^{-\frac{3 \phi}{4}}}{4} \frac{\rho}{|\kappa|} \left(A^{-1/2} + A^{1/2} \right) \bigg|_{\tilde h = \tilde h_i}\,.
\end{align}
We note that the presence of the D8-brane selects a particular gauge for $B_2$, for which the fields are continuous.

\begin{figure}[ht]
\begin{subfigure}[b]{0.5\textwidth}
\centering
\includegraphics[width=.95\linewidth]{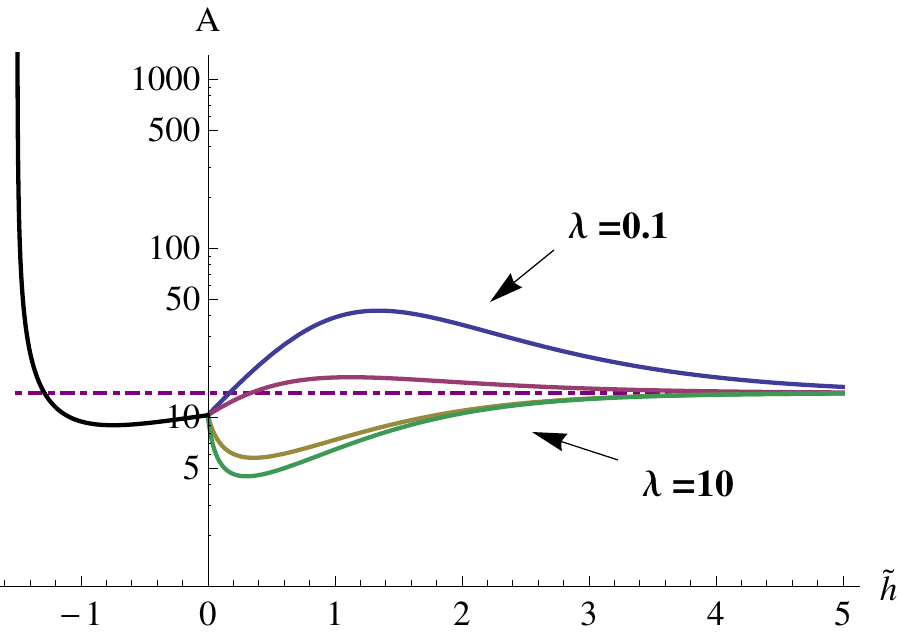}
\end{subfigure}
\begin{subfigure}[b]{0.5\textwidth}
\centering
\includegraphics[width=.95\linewidth]{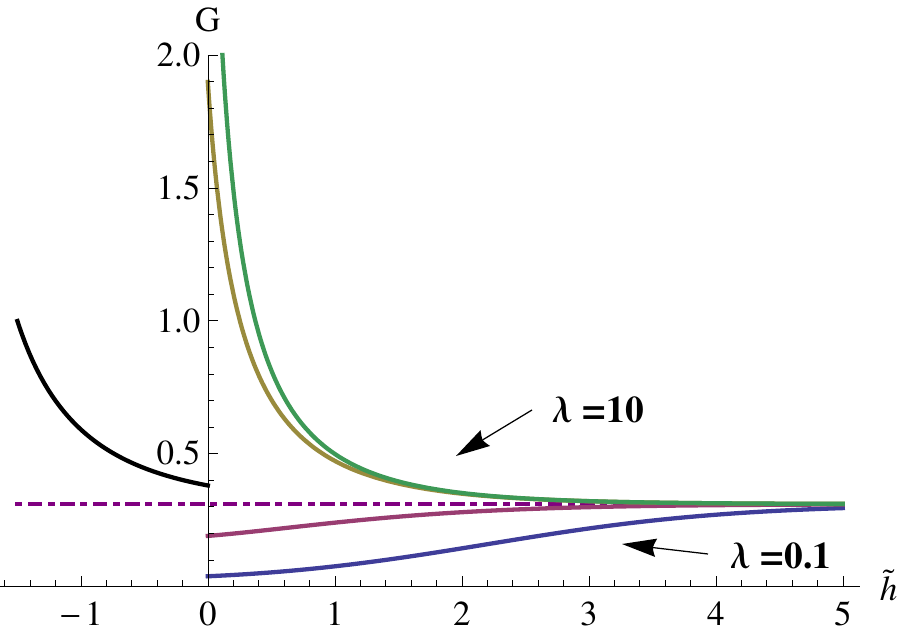}
\end{subfigure}
\caption{Plot of $A$ (left) and $G$ (right) versus $\tilde h$.  The black line is a type II solution with $G_0 = 1$.  We take $F_0$ to jump at $\tilde h = 0$ with jump coefficient $\lambda = (0.1,0,5,5,10)$.  The dot-dashed horizontal lines are the constant $G$ solution of section \ref{sec:simpsol}.}
\label{jumpp}
\end{figure}

\begin{figure}[ht]
\begin{subfigure}[b]{0.5\textwidth}
\centering
\includegraphics[width=.95\linewidth]{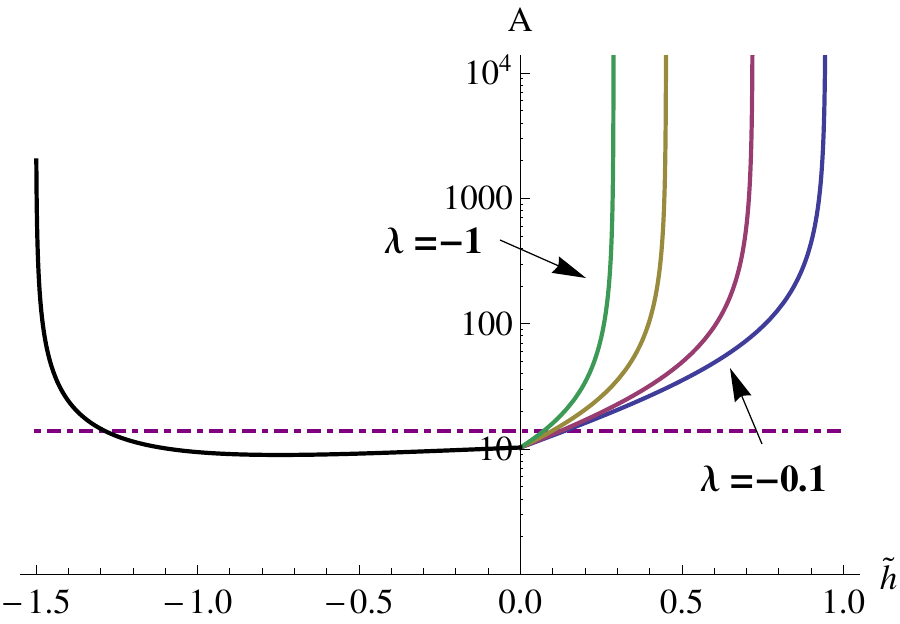}
\end{subfigure}
\begin{subfigure}[b]{0.5\textwidth}
\centering
\includegraphics[width=.95\linewidth]{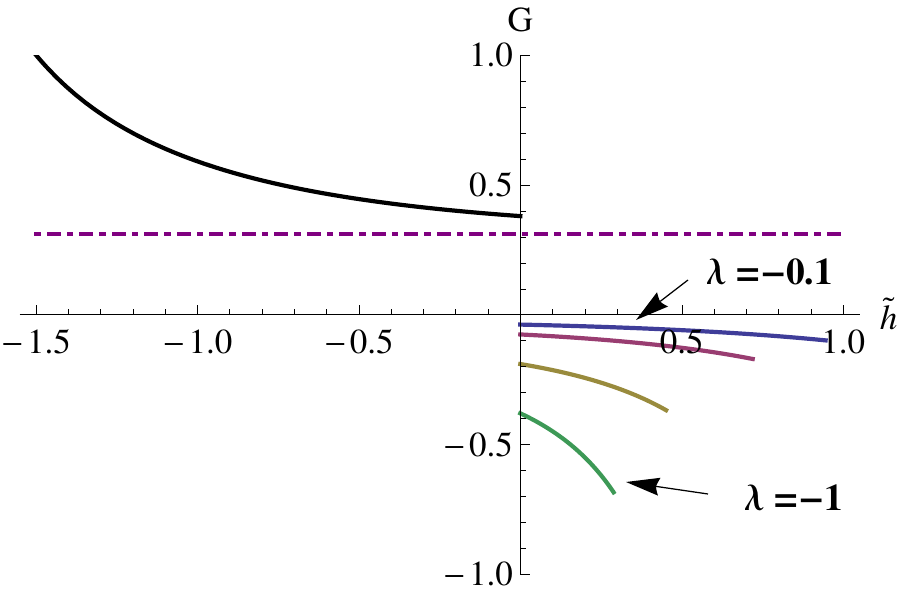}
\end{subfigure}
\caption{Plot of $A$ (left) and $G$ (right) versus $\tilde h$.  The black line is a type II solution with $G_0 = 1$.  We take $F_0$ to jump at $\tilde h = 0$ with jump coefficient $\lambda = (-0.1,-0.2,-0.5,-1)$.  The dot-dashed horizontal lines are the constant $G$ solution of section \ref{sec:simpsol}.}
\label{jumpm}
\end{figure}

An example for a type II solution with $G_0 =1$ is show in figure \ref{jumpp}.  We have chosen $\lambda > 0$ so that the value of $F_0$ is the same on either side of the D8-brane.  We see that the solution always interpolates to the constant solution for any value of the jump.  This just means that we are gluing together two type II solutions.  In figure \ref{jumpm}, we consider the same initial function but now take negative values for $\lambda$.  Note that since both $G$ and $F_0$ are negative after the jump, their product and more specifically, the sign of $\rho$, are positive.  If $\rho$ had flipped signs across the jump, it would be discontinuous.  In this case, we again find that we are gluing together two type II solutions, now one with positive $G$ and one with negative $G$.  In this case, the geometry does not approach the asymptotically flat solution of section \ref{sec:simpsol}, but rather at each end the geometry caps off as the asymptotic type II solution of \eqref{asymc1c2a}.

In general, one could also start with the type I solution or right type V solution and introduce a jump.  We encounter similar behavior with positive values of $\lambda$ yielding a geometry which asymptotes to the constant solution and negative values of $\lambda$ resulting in a type II cap.  The type III, IV and left V geometries yield reflected versions of the previous cases.

One may wonder whether we can construct solutions which interpolate from the smooth cap of the type I solution to the smooth cap of the type III solution or to the O8-caps of the type V solutions.  Unfortunately, this does not seem possible.  One necessary requirement to patch together two different solutions is for their $A$ values to overlap at some point.  Examining figure \ref{case135}, we observe that the type I, III and V solutions are disconnected, with their $A$ values never overlapping.  Thus there are no solutions which yield geometries of the form $AdS_2 \times {\cal M}_8$ with ${\cal M}_8$ compact.

\subsection{The case $F_0 =0$}

It turns out there are no solutions with $F_0 = 0$.  To see this we first take the $F_0 \rightarrow 0$ limit of the BPS equations of section \ref{sec:summBPS}.  As discussed at the end of the section, we send $b_0 \rightarrow b_0 - b_1/F_0$ and then take $F_0 = 0$.
  We solve \eqref{algconsts} to obtain expressions for $b_1$ and $b_z$:
\begin{align}
&b_1 =
- 2 e^{-3\phi/4} \frac{f_1^2}{|\alpha|^2 + |\beta|^2} \left( 2 c_1 - 3 c_2 + 3 c_3 \right)\,,&
&b_z = \frac{3}{2} \frac{e^{\phi/2}}{\alpha^* \beta^*} \left( 2 c_1 - 3 c_2 + 3 c_3 \right)
\,,&
\end{align}
along with $c_3 = - c_2$.  The differential equations \eqref{eq:diff} can be reduced as before to yield $\bar{\kappa}= \rho/\alpha \beta$,
$b_z =(2 e^{\phi/2} / \rho) \partial_z \ln \left(\alpha/\beta\right)$ and
\begin{align}
\partial_z\ln\left(\alpha(\beta^*)^2\sqrt{\rho}\right)=&
- \frac{1}{8} \frac{b_1}{f_1^2}e^{-\phi/2} e^{5\phi/4}\frac{\rho\alpha}{\beta^*}\,, \cr
\partial_z\ln\left(\beta(\alpha^*)^2\sqrt{\rho}\right)=&
\frac{1}{8} \frac{b_1}{f_1^2}e^{-\phi/2} e^{5\phi/4}\frac{\rho\beta}{\alpha^*}\,.
\end{align}
We eliminate $\beta$ using the definition of $\kappa$.  Requiring $b_1$ to be constant, we can obtain algebraic expressions for $\p_z \alpha$, $\p_z \alpha^*$, $\p_z \rho$ and $\p_z \phi$.  Upon plugging these expressions into \eqref{eq:dil}, we obtain an algebraic equation which over constrains the system.  To exhibit this explicitly, it is convenient to pick local coordinates so that $\kappa = \bar \kappa = 1/2$.  Then \eqref{eq:dil} yields
\begin{align}
4 \rho^2 + |\alpha|^4 = 0 \,,
\end{align}
which has no solutions other than $\rho = \alpha = 0$.

\section{Summary}
We studied $D(2,1;\gamma;1) \times SO(4)$ geometries in massive IIA supergravity and reduced the BPS-equations to a two-dimensional system. This two-dimensional system has three points of supersymmetry enhancement. Two of these lead to the $AdS_6$ solution of \cite{Brandhuber:1999np} which we rederived as the only solution. The remaining point of supersymmetry enhancement leads to novel solutions.

We found a solution describing a fundamental string in the $F_0$ background (constant A and G solution), a set of solutions corresponding to fundamental strings ending on D8-branes (case I and III) and another set of novel solutions corresponding to fundamental strings ending on an O8-plane (case V). We briefly discussed glueing solutions with different values of $F_0$ together using D8-brane domain walls. In all cases there naively seems to be no decoupling limit, as the geometries contain asymptotically flat regions.  However, we note that the string coupling tends to zero in these regions. The limiting case $F_0=0$ (regular 'massless' IIA) admits no solutions in these points of enhanced symmetry. Finally, we find that there are no geometries of the form $AdS_2 \times \mathcal{M}_8$ with $\mathcal{M}_8$ compact such that there are no solutions corresponding to $1+0$ dimensional CFTs.

\acknowledgments

We thank C. Bachas, S. Cremonesi, E. D'Hoker, A. Hanany, J. Gauntlett, and T. Van Riet for useful conversations and correspondence.  J.E. acknowledges the hospitality of NYCCT and its CTP where part of this research was carried out.  D.K. and B.V.P. acknowledge the hospitality of Imperial College London where part of this research was carried out. The research of D.K. was supported in part by the PSC-CUNY Research Award.\\
The work of B.V.P was supported in part by the FWO - Vlaanderen, Project No.
G.0651.11, and in part by the Interuniversity Attraction Poles Programme
initiated by the Belgian Science Policy (P7/37).

\appendix

\section{Massive IIA supergravity}
\label{app:typeIIA}

We follow the conventions of \cite{Romans:1985tz}, with the following replacements
$\phi \rightarrow -\phi/2$, $\xi \rightarrow e^{-\phi/4}$, $G_{MNP} \rightarrow H_{MNP}/2$, $F_{MNPQ} \rightarrow F_{MNPQ}/2$ and $m \rightarrow F_0$. The action is given by
\begin{align}
S_{IIA} =& \frac{1}{2 \kappa^2} \int d^{10}x \sqrt{-g} \bigg( R - \frac{1}{2} (\p_\mu \Phi)(\p^\mu \Phi) - \frac{e^{\Phi/2}}{48} F^{MNPQ} F_{MNPQ}
- \frac{e^{-\Phi}}{12} H^{MNP} H_{MNP}
\cr
& - F_0^2 \frac{e^{3 \Phi/2}}{4} B^{MN} B_{MN}
- F_0^2 \frac{e^{5\Phi/2}}{2}
 \bigg)
 + \frac{1}{4 \kappa^2} \int d^{10}x \ \bigg( d C_{3} \wedge d C_{3} \wedge B_{2}
  \cr&
  +\frac{1}{3} F_0 \ d C_{3} \wedge B_{2} \wedge B_{2} \wedge B_{2} + \frac{1}{20} F_0^2 \  B_{2} \wedge B_{2} \wedge  B_{2} \wedge B_{2} \wedge B_{2} \bigg) \,.\no
\end{align}
The field strengths are related to the gauge potentials by
\begin{align}
&H_{3} = d B_{2}\,,&
&F_{4} = d C_{3} + \frac{F_0}{2} B_{2} \wedge B_{2}\,,&
\end{align}
while $F_0$ is step-wise constant.  The field strengths satisfy the following Bianchi identities
\begin{align}
&d H_{3} = 0,&
&d F_{4} = F_0 \, B_{2} \wedge H_{3}.&
\end{align}
In IIA supergravity, the spinor satisfies the reality condition $\epsilon^* = {\cal B} \epsilon$, with ${\cal B}$ given in \eqref{defB}.  The BPS equations in Einstein frame are given by \cite{Romans:1985tz}\footnote{Compared to \cite{Romans:1985tz}, we have redefined $\lambda$ by a multiplicative factor of $2 \sqrt 2$.}
\begin{align}
\label{app:BPSromans}
\delta \lambda =& \bigg[ (D_M \phi) \Gamma^M + \frac{5}{4} F_0
 e^{\frac{5}{4} \phi} + \frac{1}{96} e^\frac{\phi}{4} (F_{MNPQ}
 \Gamma^{MNPQ})
 \cr &
 - \frac{3}{8} F_0  e^{\frac{3\phi}{4}} B_{MN} \Gamma^{MN} \Gamma_{11}
 - \frac{1}{12} e^{-\frac{\phi}{2}} H_{MNP} \Gamma^{MNP} \Gamma_{11}
 \bigg] \epsilon = 0 \ , \\
\delta \psi_M =& \bigg[ D_M - \frac{1}{32} F_0 e^{\frac{5}{4} \phi}
 \Gamma_M + \frac{1}{128} \frac{e^{\frac{\phi}{4}}}{2} F_{NPQR} (
 \Gamma_M {}^{NPQR} - \frac{20}{3} \delta_M {}^N \Gamma^{PQR} )
 \cr&
 - \frac{1}{32} F_0 \frac{e^{\frac{3\phi}{4}}}{2} B_{NP} (\Gamma_M {}^{NP} - 14 \delta_M {}^N \Gamma^P) \Gamma_{11}
 \cr&
 + \frac{1}{48} \frac{e^{-\frac{\phi}{2}}}{2} H_{NPQ} (\Gamma_M {}^{NPQ} - 9 \delta_M {}^N \Gamma^{PQ}) \Gamma_{11}
 \bigg]
 \epsilon = 0 \ . \no
\end{align}
In this formulation, one cannot explicitly take the massless limit, $F_0 \rightarrow 0$.  However, the massless IIA limit can be recovered by first making the gauge transformation $B_{2} \rightarrow B_{2} - F_0^{-1} F_{2}$ with $d F_{2} = 0$ and $C_{3} \rightarrow C_{3} - (2 F_0)^{-1} A_{1} \wedge F_{2}$.  Note that $F_{4}$ is invariant under these combined transformations.  One can then take $F_0 \rightarrow 0$.

Massive IIA supergravity coupled to D8-branes and O8-planes was formulated in \cite{Bergshoeff:2001pv}.  We denote their quantities with hats.  In their notation, the BPS equations are given in string frame by
\begin{align}
\delta \hat \lambda =& \left( \Gamma^M (D_M \phi) + \frac{1}{12} H_{MNP} \Gamma^{MNP} \Gamma_{11} \right) \epsilon \cr
& + \frac{1}{4} e^\phi
\left(5 G^{(0)} + \frac{3}{2} G_{MN}^{(2)} \Gamma^{MN} \Gamma_{11} + \frac{1}{24} G^{(4)}_{MNPQ} \Gamma^{MNPQ} \right) \epsilon\,, \cr
\delta \hat \psi_M =& (D_M + \frac{1}{8} \Gamma_{11} \hat H_{MNP} \Gamma^{NP}) \epsilon
\cr &
+ \frac{1}{8} e^\phi \left( G^{(0)} \Gamma_M + \frac{1}{2} G^{(2)}_{NP} \Gamma^{NP} \Gamma_M \Gamma_{11} + \frac{1}{24} G^{(4)}_{NPQR} \Gamma^{NPQR} \Gamma_M \right) \epsilon \,,
\end{align}
where the fluxes $G^{(2n)}$ are given by
\begin{align}
&G^{(0)} = F_0\,, \qquad
G^{(2)} = d \hat A_{1} + F_0 \hat B_{2} \,, \cr
&G^{(4)} = d \hat C_{3} + d \hat A_{1} \wedge \hat B_{2} + \frac{F_0}{2} \hat B_{2} \wedge \hat B_{2} \,.
\end{align}
with $\hat A_{1}$ a $1$-form and $\hat C_{3}$ a $3$-form and $F_0$ is the Romans mass.  To compare to \eqref{app:BPSromans}, we first go to Einstein frame, related to string frame by $ds_{string}^2 = e^{\phi/2} ds^2_{Einstein}$, and substitute in for the $G^{(2n)}$.
We introduce  $\hat F_{4} = d \hat C_{3} + (F_0/2) \hat B_{2} \wedge \hat B_{2}$.  After transforming the above equations to Einstein frame, they can be mapped to \eqref{app:BPSromans} by the following field redefinitions\footnote{Upon going to Einstein frame, the dilaton enters into the expression for $\delta \hat \psi_M$.  To eliminate this contribution, we introduce $\psi_M = e^{\phi/2} \hat \psi_M - (\Gamma_M/8) e^{\phi/2} \hat \lambda$ and $\lambda = e^{\phi/2} \hat \lambda$.}
\begin{align}
\label{app:masslesstrans}
\hat B_{2} =& -B_{2} - \frac{1}{F_0} \hat F_{2}\,, \cr
\hat C_{3} =& C_{3} + \frac{1}{2 F_0} \hat A_{1} \wedge \hat F_{2}\,.
\end{align}

\section{Gamma matrix conventions}
\label{app:gamma}
We choose a basis for the Clifford algebra which is well-adapted to the $AdS_{2} \times S^3 \times S^3 \times \Sigma_2$ space.
\begin{align}
\Gamma^{m} &= \gamma^{m}\otimes I_2 \otimes I_2 \otimes I_4& m&=0,1
\no \\
\Gamma^{i} &= \sigma^3 \otimes \gamma^{i} \otimes I_2 \otimes \left(\sigma^1\otimes I_2 \right)& i&=2,3,4
\no \\
\Gamma^{\tilde{i}} &= \sigma^3 \otimes I_2 \otimes \gamma^{\tilde{i}} \otimes \left(\sigma^2\otimes I_2 \right)& \tilde{i}&=5,6,7
\no \\
\Gamma^{a} &= \sigma^3 \otimes I_2 \otimes I_2 \otimes \gamma^a& a&=8,9\,,
\end{align}
where a convenient basis for the lower Clifford algebras is as follows,
\begin{align}
i\gamma^{0} = \sigma^1 &=\gamma^2 = \gamma^5  & \gamma^8&=\sigma^3\otimes\sigma^1
\no \\
\gamma^{1} = \sigma^2 &= \gamma^3 =\gamma^6 & \gamma^9&=\sigma^3\otimes\sigma^2
\no \\
\sigma^3 &= \gamma^4 =\gamma^7\,. & &
\end{align}
The $10$-dimensional chirality matrix in this basis is given by
\begin{align}
\Gamma^{11} = \Gamma^{0123456789} = \sigma^3 \otimes I_2 \otimes I_2 \otimes \sigma^3 \otimes \sigma^3\,.
\end{align}
The complex conjugation matrices in each subspace are defined by
\begin{align}
(\gamma^m)^* = + B_{(1)}\gamma^m B_{(1)}^{-1} \qquad (B_{(1)})^*B_{(1)} &= + I_2 \qquad B_{(1)}=\sigma^3
\no \\
(\gamma^i)^* = - B_{(2)}\gamma^i B_{(2)}^{-1} \qquad (B_{(2)})^*B_{(2)} &= - I_2 \qquad B_{(2)}=\sigma^2
\no \\
(\gamma^{\tilde{i}})^* = - B_{(3)}\gamma^{\tilde{i}} B_{(3)}^{-1} \qquad (B_{(3)})^*B_{(3)} &= - I_2 \qquad B_{(3)}=\sigma^2
\no \\
(\gamma^a)^* = + B_{(4)}\gamma^a B_{(4)}^{-1} \qquad (B_{(4)})^*B_{(4)} &= + I_2 \qquad B_{(4)}=\sigma^3\otimes\sigma^1\,,
\end{align}
where in the last column we have also listed the form of these matrices in our particular basis. The $10$-dimensional complex conjugation matrix $\cB$ is defined by $(\Gamma^M)^*=\cB\Gamma^M\cB^{-1}$ and $\cB\cB^*=1$, and in this basis is given by
\begin{align}
\label{defB}
\cB=B_{(1)} \otimes B_{(2)} \otimes B_{(3)} \otimes \sigma^2 \otimes \sigma^2 = \sigma^3 \otimes \sigma^2 \otimes \sigma^2\otimes \sigma^2\otimes \sigma^2\,.
\end{align}

\section{Reduction of BPS equations to two dimensions}
\label{app:red1}

We introduce complex frames $e^z$ for the components along $\Sigma_2$, defined so that $p_z = p_8 -  i p_9$ and
\beq
p_a\sigma^a=\left(
              \begin{array}{cc}
                0 & p_z \\
                p_{\bar{z}} & 0 \\
              \end{array}
            \right)\,.
\eeq
We begin by reducing the first BPS equation given in \eqref{BPSeqns}:
\bea
\label{eA:dil}
0&=\left(\frac{1}{2\sqrt{2}}D_M\phi\Gamma^M+\frac{5}{8\sqrt{2}}F_{0}e^{5\phi/4}+\frac{1}{192\sqrt{2}}e^{\phi/4}F_{MNPQ}\Gamma^{MNPQ}\right.\nonumber \\
&\left.-\frac{3}{16\sqrt{2}}F_{0}e^{3\phi/4}B_{MN}\Gamma^{MN}\Gamma_{11}-\frac{1}{24\sqrt{2}}e^{-\phi/2}H_{MNP}\Gamma^{MNP}\Gamma_{11}\right)\epsilon \,.
\eea
The first term is decomposed as
\begin{align}
&D_M\phi\Gamma^M \epsilon = D_a \phi \Gamma^a \epsilon \cr
&= D_a \phi (\sigma^3 \otimes I_2 \otimes I_2 \otimes \gamma^a) \sum_{\eta_1,\eta_2,\eta_3} \chi_{\eta_1,\eta_2,\eta_3} \otimes \left[ \zeta_{\eta_1,\eta_3,\eta_3} \otimes \left(
                                                                                                                  \begin{array}{c}
                                                                                                                    1 \\
                                                                                                                    0 \\
                                                                                                                  \end{array}
                                                                                                                \right)
+  \hat \zeta_{\eta_1,\eta_3,\eta_3} \otimes \left(
                                                                                                                  \begin{array}{c}
                                                                                                                    0 \\
                                                                                                                    1 \\
                                                                                                                  \end{array}
                                                                                                                \right)
\right] \cr
&=
\sum_{\eta_1,\eta_2,\eta_3} \chi_{\eta_1,\eta_2,\eta_3} \otimes \left[ \sigma^3 \zeta_{-\eta_1,\eta_3,\eta_3} \otimes \left(
                                                                                                                  \begin{array}{c}
                                                                                                                    D_{\bar z} \phi \\
                                                                                                                    0 \\
                                                                                                                  \end{array}
                                                                                                                \right)
+  \sigma^3 \hat \zeta_{-\eta_1,\eta_3,\eta_3} \otimes \left(
                                                                                                                  \begin{array}{c}
                                                                                                                    0 \\
                                                                                                                    D_z \phi \\
                                                                                                                  \end{array}
                                                                                                                \right)
\right]
\cr
&=
\sum_{\eta_1,\eta_2,\eta_3} \chi_{\eta_1,\eta_2,\eta_3} \otimes \left[ \tau^{(100)} \sigma^3 \zeta_{\eta_1,\eta_3,\eta_3} \otimes \left(
                                                                                                                  \begin{array}{c}
                                                                                                                    D_{\bar z} \phi \\
                                                                                                                    0 \\
                                                                                                                  \end{array}
                                                                                                                \right)
+  \tau^{(100)} \sigma^3 \hat \zeta_{\eta_1,\eta_3,\eta_3} \otimes \left(
                                                                                                                  \begin{array}{c}
                                                                                                                    0 \\
                                                                                                                    D_z \phi \\
                                                                                                                  \end{array}
                                                                                                                \right)
\right]\,,
\end{align}
where we have written the expression so that $\chi_{\eta_1,\eta_2,\eta_3}$ is an overall coefficient.  In the last line, we have used the $\tau^{(ijk)}$ notation introduced in section \ref{sec:BPSeqns}.  Proceeding in a similar manner for each term and requiring \eqref{eA:dil} to hold independently for each $\chi_{\eta_1,\eta_2,\eta_3}$ yields two equations
\begin{align}
\label{app1:dil}
0 =& \frac{1}{2\sqrt{2}}D_z \phi \tau^{(100)} \sigma^3 \hat{\zeta}+\frac{5}{8\sqrt{2}}F_{0}e^{5\phi/4}\zeta +\frac{1}{8\sqrt{2}}e^{\phi/4}(ih_1 \tau^{(100)}\zeta +h_z\sigma^2\hat{\zeta}-g_z\sigma^1\hat{\zeta}) \nonumber\\
 &- \frac{3}{8\sqrt{2}}F_{0}e^{3\phi/4}\left(\frac{b_0}{f_1^2}\sigma^3\zeta+ i b_1 \tau^{(100)}\sigma^3\zeta\right)+\frac{1}{4\sqrt{2}}e^{-\phi/2} \frac{D_{z}b_0}{f_1^2} \tau^{(100)} \hat{\zeta}\,, \cr
0=&\frac{1}{2\sqrt{2}}D_{\bar{z}} \phi \tau^{(100)} \sigma^3 \zeta+\frac{5}{8\sqrt{2}}F_{0}e^{5\phi/4}\hat{\zeta} +\frac{1}{8\sqrt{2}}e^{\phi/4}(-ih_1 \tau^{(100)}\hat{\zeta} +h_{\bar{z}}\sigma^2\zeta-g_{\bar{z}}\sigma^1 \zeta)\nonumber\\
&  - \frac{3}{8\sqrt{2}}F_{0}e^{3\phi/4}\left(-\frac{b_0}{f_1^2}\sigma^3\hat{\zeta}+ i b_1 \tau^{(100)}\sigma^3\hat{\zeta}\right)-\frac{1}{4\sqrt{2}}e^{-\phi/2} \frac{D_{\bar{z}}b_0}{f_1^2} \tau^{(100)} \zeta\,.
\end{align}
Next we use the type IIA reality condition $\hat \zeta^* = i \sigma_2 \zeta$ to eliminate $\hat \zeta$ from the above two equations.  After doing so, we find that the equations are complex conjugates of each other, thus it is sufficient to keep only one.  As a result, \eqref{eA:dil} is equivalent to the single equation
\begin{align}
\label{app1:dil}
0&=\frac{1}{2\sqrt{2}}D_z \phi \tau^{(100)} \sigma^3 \zeta^* + \frac{5}{8\sqrt{2}}F_{0}e^{5\phi/4} i \sigma^2 \zeta -\frac{1}{8\sqrt{2}}e^{\phi/4}(h_1 \tau^{(100)} \sigma^2 \zeta + h_z \sigma^2 \zeta^*  + g_z \sigma^1 \zeta^* ) \nonumber\\
&+ \frac{3}{8\sqrt{2}}F_{0}e^{3\phi/4}\left(\frac{b_0}{f_1^2}\sigma^1\zeta + i b_1 \tau^{(100)}\sigma^1\zeta\right)-\frac{1}{4\sqrt{2}}e^{-\phi/2} \frac{D_{z}b_0}{f_1^2} \tau^{(100)} \zeta^*\,.
\end{align}

We proceed in a similar manner for the second equation of \eqref{BPSeqns}.  Note in general that this equation is differential in the spinor parameter $\epsilon$.  First we consider when the index $M$ is along the symmetric spaces.  In this case, the equation reduces to an algebraic equation in $\epsilon$.  This is achieved by making use of the Killing spinor equations for the $\chi_{\eta_1,\eta_2,\eta_3}$
\begin{align}
\label{eq:KS}
0 =& \left( \hat e^\mu_m \hat \nabla_\mu - \frac{\eta_1}{2} \gamma_m \otimes I_2 \otimes I_2  \right) \chi_{\eta_1,\eta_2,\eta_3}\,, \cr
0 =& \left( \hat e^\mu_{i} \hat \nabla_{\mu} - i \frac{\eta_2}{2} I_2 \otimes \gamma_i \otimes I_2  \right) \chi_{\eta_1,\eta_2,\eta_3}\,, \cr
0 =& \left( \hat e^\mu_{\tilde i} \hat \nabla_{\mu} - i \frac{\eta_3}{2} I_2 \otimes I_2 \otimes \gamma_{\tilde i} \right) \chi_{\eta_1,\eta_2,\eta_3}\,,
\end{align}
and noting that $\zeta$ and $\hat \zeta$ only depend on $\Sigma_2$.  We use the relation between $\zeta$ and $\hat \zeta$ to rewrite the equations in terms of $\zeta$.  The result is a reduction to a system of three equations, one for each symmetric space,
\begin{align}
\label{app1:grav1}
0=& -\frac{1}{2f_1}\tau^{(300)}\zeta+\frac{1}{2}D_{z}\ln f_{1} \tau^{(100)} \sigma^1 \zeta^* - \frac{1}{32}F_{0}e^{5\phi/4}\zeta \nonumber\\
&+\frac{3}{16}\frac{e^{\phi/4}}{2}\left(i h_{z} \zeta^* + g_z \sigma^3 \zeta^* -i\frac{5}{3}h_{1}\tau^{(100)}\zeta\right)
-\frac{1}{16}F_{0}\frac{e^{3\phi/4}}{2}\left[ib_1\sigma^3\tau^{(100)}\zeta-7\frac{b_{0}}{f_{1}^{2}}\sigma^3\zeta\right]\nonumber \\
&+\frac{3}{8}\frac{e^{-\phi/2}}{2}\frac{D_z b_0}{f_1^2}\tau^{(100)} i \sigma^2 \zeta^*\,,\\
\label{app1:grav2}
0=&
-\frac{i}{2f_2}\sigma_{1}\tau^{(030)}\zeta+\frac{1}{2}D_{z}\ln f_{2} \tau^{(100)} \sigma^1 \zeta^* - \frac{1}{32}F_{0}e^{5\phi/4}\zeta\nonumber\\
&+\frac{3}{16}\frac{e^{\phi/4}}{2}\left(i h_{1} \tau^{(100)}\zeta+ g_z \sigma^3 \zeta^*-i \frac{5}{3}h_{z} \zeta^* \right)
-\frac{1}{16}F_{0}\frac{e^{3\phi/4}}{2}\left[\frac{b_{0}}{f_{1}^{2}}\sigma^3\zeta+ib_1\sigma^3\tau^{(100)}\zeta\right]\nonumber\\
&-\frac{1}{8}\frac{e^{-\phi/2}}{2}\frac{D_z b_0}{f_1^2}\tau^{(100)} i\sigma^2 \zeta^*\,,
\end{align}
\begin{align}
\label{app1:grav3}
0=&
-\frac{i}{2f_3}\sigma_{2}\tau^{(003)}\zeta+\frac{1}{2}D_{z}\ln f_{3} \tau^{(100)} \sigma^1 \zeta^* - \frac{1}{32}F_{0}e^{5\phi/4}\zeta\nonumber\\
&+\frac{3}{16}\frac{e^{\phi/4}}{2}\left(i h_{1} \tau^{(100)}\zeta+ i h_z \zeta^* -\frac{5}{3}g_{z} \sigma^3 \zeta^*\right)
-\frac{1}{16}F_{0}\frac{e^{3\phi/4}}{2}\left[\frac{b_{0}}{f_{1}^{2}}\sigma^3\zeta+i b_1\sigma^3\tau^{(100)}\zeta\right]\nonumber\\
&-\frac{1}{8}\frac{e^{-\phi/2}}{2}\frac{D_z b_0}{f_1^2}\tau^{(100)} i \sigma^2 \zeta^*\,.
\end{align}
Finally, we reduce \eqref{BPSeqns} for the components along $\Sigma_2$.  In this case, the equations remain differential in $\epsilon$ and are given by
\begin{align}
\label{app1:grav4}
0=&D_{z}\zeta - \frac{1}{2} D_z(\ln \rho) \,  \zeta
+\frac{3}{16}\frac{e^{\phi/4}}{2}\left[i \frac{8}{3} h_z \sigma^1 \tau^{(100)} \zeta + i \frac{8}{3} g_z \sigma^2  \tau^{(100)} \zeta  \right]
-\frac{1}{2}\frac{e^{-\phi/2}}{2}\frac{D_{z} b_0}{f_1^2} \sigma^3 \zeta\,,\\
\label{app1:grav5}
0=&D_{\bar z} \zeta + \frac{1}{2} D_{\bar z} (\ln \rho) \zeta - \frac{1}{16}F_{0}e^{5\phi/4} \sigma^1 \tau^{(100)} \zeta^*
+\frac{3}{16}\frac{e^{\phi/4}}{2}\left[i \frac{2}{3} h_{\bar z} \sigma^1 \tau^{(100)} \zeta + i \frac{2}{3} g_{\bar z} \sigma^2 \tau^{(100)}\zeta \right.\nonumber\\&
\left. + i\frac{10}{3} h_1 \sigma^1 \zeta^*\right]
+\frac{1}{8}F_{0}\frac{e^{3\phi/4}}{2}\left[\frac{b_{0}}{f_{1}^{2}} \tau^{(100)} i \sigma^2 \zeta^* - 7 b_1 \sigma^2 \zeta^* \right]
-\frac{1}{4}\frac{e^{-\phi/2}}{2}\frac{D_{\bar z} b_0}{f_1^2} \sigma^3\zeta\,.
\end{align}

\section{The general case}
\label{app:gencase}

In the case when $b_1$ or $h_1$ no longer vanish, $\tau^{(300)} \sigma^3$ no longer commutes with the BPS equations.  It is still useful to decompose the BPS equations in terms of eigenstates of $\tau^{(300)} \sigma^3$.  We denote the two eigenstates by $\xi_{\pm}$.
The equations then reduce to
\begin{align}
0
&=\frac{1}{2\sqrt{2}}D_z \phi \sigma^3 \xi_{\pm}^* + \frac{5}{8\sqrt{2}}F_{0}e^{5\phi/4} i \sigma^2 \xi_{\pm} -\frac{1}{8\sqrt{2}}e^{\phi/4}(h_z \sigma^2 \xi_{\pm}^*  + g_z \sigma^1 \xi_{\pm}^* ) \nonumber\\
&+ \frac{3}{8\sqrt{2}}F_{0}e^{3\phi/4} \frac{b_0}{f_1^2} \sigma^1 \xi_{\pm} - \frac{1}{4\sqrt{2}}e^{-\phi/2} \frac{D_{z}b_0}{f_1^2} \xi_{\pm}^*
\cr
&- \frac{1}{8\sqrt{2}} e^{\phi/4} h_1 \sigma^2 \xi_{\mp} + \frac{3}{8 \sqrt{2}} F_0 e^{3 \phi/4} i b_1 \sigma^1 \xi_{\mp}\,.
\end{align}
The gravitino equation reduces to the algebraic equations
\begin{align}
0=& -\frac{(\pm 1)}{2f_1} \sigma^3 \xi_{\pm}+\frac{1}{2}D_{z}\ln f_{1} \sigma^1 \xi_{\pm}^* - \frac{1}{32}F_{0}e^{5\phi/4}\xi_{\pm} + \frac{3}{16}\frac{e^{\phi/4}}{2}\left(i h_{z} \xi_{\pm}^* + g_z \sigma^3 \xi_{\pm}^* \right)
\nonumber\\ &
+ \frac{7}{16}F_{0}\frac{e^{3\phi/4}}{2} \frac{b_{0}}{f_{1}^{2}}\sigma^3\xi_{\pm}
+\frac{3}{8}\frac{e^{-\phi/2}}{2}\frac{D_z b_0}{f_1^2} i \sigma^2 \xi_{\pm}^*
- \frac{5}{16}\frac{e^{\phi/4}}{2} i h_{1} \xi_{\mp}
-\frac{1}{16}F_{0}\frac{e^{3\phi/4}}{2} i b_1 \sigma^3 \xi_{\mp}
\,,\cr
0=&
-\frac{i \nu_2}{2f_2}\sigma_{1} \xi_{\pm} +\frac{1}{2}D_{z}\ln f_{2} \sigma^1 \xi_{\pm}^* - \frac{1}{32}F_{0}e^{5\phi/4}\xi_{\pm}
+\frac{3}{16}\frac{e^{\phi/4}}{2}\left(g_z \sigma^3 \xi_{\pm}^*-i \frac{5}{3}h_{z} \xi_{\pm}^* \right)
\nonumber\\ &
-\frac{1}{16}F_{0}\frac{e^{3\phi/4}}{2} \frac{b_{0}}{f_{1}^{2}}\sigma^3\xi_{\pm}
-\frac{1}{8}\frac{e^{-\phi/2}}{2}\frac{D_z b_0}{f_1^2}  i\sigma^2 \xi_{\pm}^*
+\frac{3}{16}\frac{e^{\phi/4}}{2} i h_{1} \xi_{\mp} -\frac{1}{16}F_{0}\frac{e^{3\phi/4}}{2} i b_1 \sigma^3 \xi_{\mp}
\,,\cr
0=&
-\frac{i \nu_3}{2f_3}\sigma_{2} \xi_{\pm}+\frac{1}{2}D_{z}\ln f_{3} \sigma^1 \xi_{\pm}^* - \frac{1}{32}F_{0}e^{5\phi/4}\xi_{\pm} +\frac{3}{16}\frac{e^{\phi/4}}{2}\left(i h_z \xi_{\pm}^* -\frac{5}{3}g_{z} \sigma^3 \xi_{\pm}^*\right)
\nonumber\\&
-\frac{1}{16}F_{0}\frac{e^{3\phi/4}}{2} \frac{b_{0}}{f_{1}^{2}}\sigma^3\xi_{\pm}
-\frac{1}{8}\frac{e^{-\phi/2}}{2}\frac{D_z b_0}{f_1^2} i \sigma^2 \xi_{\pm}^*
+\frac{3}{16}\frac{e^{\phi/4}}{2} i h_{1} \xi_{\mp}
-\frac{1}{16}F_{0}\frac{e^{3\phi/4}}{2} i b_1\sigma^3 \xi_{\mp}\,,
\end{align}
and the differential equations
\begin{align}
0=&D_{z}\xi_{\pm} - \frac{1}{2} D_z(\ln \rho) \,  \xi_{\pm}
+\frac{3}{16}\frac{e^{\phi/4}}{2}\left[i \frac{8}{3} h_z \sigma^1 \xi_{\pm} + i \frac{8}{3} g_z \sigma^2  \xi_{\pm} \right]
-\frac{1}{2}\frac{e^{-\phi/2}}{2}\frac{D_{z} b_0}{f_1^2} \sigma^3 \xi_{\pm}\,, \cr
0=&D_{\bar z} \xi_{\pm} + \frac{1}{2} D_{\bar z} (\ln \rho) \xi_{\pm} - \frac{1}{16}F_{0}e^{5\phi/4} \sigma^1 \xi_{\pm}^*
+\frac{3}{16}\frac{e^{\phi/4}}{2}\left[i \frac{2}{3} h_{\bar z} \sigma^1 \xi_{\pm} + i \frac{2}{3} g_{\bar z} \sigma^2 \xi_{\pm} \right]
\cr&
+i \frac{1}{8}F_{0}\frac{e^{3\phi/4}}{2} \frac{b_{0}}{f_{1}^{2}} \sigma^2 \xi_{\pm}^*
-\frac{1}{4}\frac{e^{-\phi/2}}{2}\frac{D_{\bar z} b_0}{f_1^2} \sigma^3 \xi_{\pm}
\cr&
+ \frac{5}{8} \frac{e^{\phi/4}}{2} i h_1 \sigma^1 \xi_{\mp}^*
-\frac{7}{8} F_{0}\frac{e^{3\phi/4}}{2} b_1 \sigma^2 \xi_{\mp}^*\,.
\end{align}

\section{Decoupling the BPS equations}
\label{app:red2}

In this appendix we further reduce the BPS equations \eqref{eq:odil}, \eqref{eq:oalg} and \eqref{eq:odiff}.  We obtain algebraic expressions for the metric factors $f_i$.  We obtain an algebraic expression for $b_0$ and show that no further constraints arise from differentiating this expression.
This allows us to treat $b_0$ and $D_z b_0$ as independent variables, since the BPS equations will enforce the differential relation among them.  We find it convenient to introduce the notation $b_z = D_z b_0/f_1^2$ and treat $b_0$ and $b_z$ as independent.  Finally we derive an algebraic constraint relating the $c_i$.  The net result is the reduction of \eqref{eq:odil} and \eqref{eq:oalg} to the expressions for the metric factors \eqref{app:met}, the four algebraic equations \eqref{solb0}, \eqref{const1}, \eqref{const2} and \eqref{algconsst} and a differential equation for the dilaton.

To solve for the metric factors, we first use \eqref{eq:odiff} to compute the following derivatives of spinor bilinears
\begin{align}
\label{eq:bidiffs}
D_z (\xi^\dagger \xi)=&
 \frac{1}{16}F_{0}e^{5\phi/4} \xi^t \sigma^1 \xi
-\frac{3}{8}\frac{e^{\phi/4}}{2}\left[i h_z \xi^\dagger \sigma^1 \xi + i g_z \xi^\dagger \sigma^2  \xi \right]
+\frac{3}{4}\frac{e^{-\phi/2}}{2}\frac{D_{z} b_0}{f_1^2} \xi^\dagger \sigma^3 \xi \,,\no\\
D_z (\xi^\dagger \sigma^2\xi)=&
 \frac{i}{16}F_{0}e^{5\phi/4} \xi^t \sigma^3 \xi
-\frac{3}{8}\frac{e^{\phi/4}}{2}\left[\frac{5}{3} h_z \xi^\dagger \sigma^3 \xi + i g_z \xi^\dagger\xi \right] \nonumber\\& \qquad
+ \frac{i}{8} F_{0}\frac{e^{3\phi/4}}{2}\frac{b_0}{f_1^2}\xi^t\xi
+\frac{i}{4}\frac{e^{-\phi/2}}{2}\frac{D_{z} b_0}{f_1^2} \xi^\dagger \sigma^1 \xi \,,\no\\
D_z (\xi^\dagger \sigma^1\xi)=&
 \frac{1}{16}F_{0}e^{5\phi/4} \xi^t \xi
-\frac{3}{8}\frac{e^{\phi/4}}{2}\left[i h_z \xi^\dagger \xi - \frac{5}{3} g_z \xi^\dagger\sigma^3\xi \right] + \frac{1}{8} F_{0}\frac{e^{3\phi/4}}{2}\frac{b_0}{f_1^2}\xi^t\sigma^3\xi \nonumber\\& \qquad
-\frac{i}{4}\frac{e^{-\phi/2}}{2}\frac{D_{z} b_0}{f_1^2} \xi^\dagger \sigma^2 \xi \,.
\end{align}
To obtain equations involving only the metric factors, we multiply the three equations given in \eqref{eq:oalg} respectively by $\xi^t \sigma^1$, $-i\xi^t\sigma^3$ and $\xi^t$.  Combining the resulting equations with the above equations yields
\begin{align}
D_z (\xi^\dagger \xi) =& D_{z}( \ln f_{1} ) \xi^t \xi^*\,, \no\\
D_z (\xi^\dagger \sigma^2 \xi) =& D_{z} (\ln f_{2}) \xi^\dagger \sigma^2 \xi\,, \no\\
D_z (\xi^\dagger \sigma^1 \xi) =& D_{z} (\ln f_{3}) \xi^\dagger \sigma^1 \xi\,.
\end{align}
These equations are integrated to give
\begin{align}
\label{app:met}
&f_1 = \frac{\nu_1}{c_1} \xi^\dagger \xi\,,&
&f_2 = \frac{\nu_2}{c_2} \xi^\dagger \sigma^2 \xi\,,&
&f_3 = \frac{\nu_3}{c_3} \xi^\dagger \sigma^1 \xi\,,&
\end{align}
where the $c_i$ are real constants.  The factors of $\nu_i$ have been introduced for convenience.

We now move on to solving algebraically for $b_0$.  We start by combining the first equation of \eqref{eq:oalg} with three-halves of the second and third equations.  Multiplying the resulting equation by $\xi^\dagger \sigma^3$ gives
\bea
0=
-\frac{\nu_1}{2f_1} \xi^\dagger \xi + \frac{3}{2} \frac{\nu_2}{2f_2} \xi^\dagger \sigma^2 \xi - \frac{3}{2} \frac{\nu_3}{2f_3} \xi^\dagger \sigma^1 \xi
- \frac{1}{8}F_{0}e^{5\phi/4} \xi^\dagger \sigma^3 \xi + \frac{1}{4}F_{0}\frac{e^{3\phi/4}}{2} \frac{b_{0}}{f_{1}^{2}} \xi^\dagger \xi\,.
\eea
Using the above expressions for the metric factors, we solve this equation for $b_0$
\begin{align}
\label{solb0}
b_0 = \frac{8\xi^\dagger \xi}{F_{0} c_1^2}e^{-3\phi/4}\left(\frac{1}{2} c_1 - \frac{3}{4} c_2 + \frac{3}{4} c_3
+ \frac{1}{8}F_{0}e^{5\phi/4} \xi^\dagger \sigma^3 \xi\right)\,.
\end{align}

Next we show that the derivative of $b_0$ is automatically reproduced by the BPS equations.  To do so, we first differentiate the above expression
\begin{align}
\label{Dzb0eq}
D_z b_0 =& \frac{8}{F_{0}c_1^2}D_z \left(\xi^{\dagger}\xi e^{-3\phi/4}\right)\left(\frac{1}{2} c_1 - \frac{3}{4} c_2 + \frac{3}{4} c_3
+ \frac{1}{8}F_{0}e^{5\phi/4} \xi^\dagger \sigma^3 \xi \right)
\nonumber\\&
+\frac{\xi^\dagger \xi}{c_1^2} e^{-3\phi/4} D_z \left(e^{5\phi/4}\xi^{\dagger}\sigma^3\xi\right)\,.
\end{align}
Next we show that the BPS equations imply this equation is automatic.  We first use \eqref{eq:odil} and \eqref{eq:odiff} to compute
\begin{align}
D_z\left(e^{5\phi/4}\xi^{\dagger}\sigma^3\xi\right) =& e^{3\phi/4}\frac{D_z b_0}{f_1^2}\xi^{\dagger}\xi -F_{0}e^{2\phi}\frac{b_0}{f_1^{2}}\xi^t\sigma^1\xi\,, \nonumber\\
D_z(\xi^\dagger \xi e^{-3 \phi/4}) =& F_{0} e^{\phi/2} \xi^t \sigma^1 \xi\,.
\end{align}
Using these expressions to eliminate $D_z\left(e^{5\phi/4}\xi^{\dagger}\sigma^3\xi\right)$ and $D_z(\xi^\dagger \xi e^{-3 \phi/4})$, as well as the first equation of \eqref{eq:bidiffs} and the expression \eqref{solb0} for $b_0$, we find that \eqref{Dzb0eq} is automatic.  As a result, we may introduce $b_z$ as $b_z = D_z b_0/f_1^2$ and treat $b_z$ and $b_0$ as independent variables.  The BPS equations will correctly enforce the relation between the two variables.

In total the system of equations \eqref{eq:odil} and \eqref{eq:oalg} provides $7$ algebraic equations and one equation differential in $\phi$.  Three of these equations are used to solve for the metric factors and a fourth equation gives $b_0$.  We exhibit the remaining three equations as follows.  We first obtain a simple equation relating $g_z$ and $h_z$.  To do so, we first take the difference of the second and third equations appearing in \eqref{eq:oalg} and multiply the resulting expression by $\xi^{\dagger}\sigma^3$ to obtain
\beq
\label{const1}
0=2\left(c_2 + c_3 \right)
+ e^{\phi/4} g_z \xi^{\dagger} \xi^* - i e^{\phi/4}  h_z \xi^{\dagger} \sigma^3 \xi^*\,.
\eeq

Next, we use the BPS equations to obtain an algebraic constraint amongst the $c_i$.  To do so, we multiply the first equation of \eqref{eq:oalg} by $\xi^\dagger \sigma^3$ and use \eqref{solb0} to eliminate $b_0$
\begin{align}
0 = &- \frac{1}{2} c_1 + \frac{3}{16} F_{0} e^{5\phi/4} \xi^\dagger \sigma^3 \xi + \frac{3}{32} e^{\phi/4} \left( g_z \xi^\dagger \xi^* + i h_z \xi^\dagger \sigma^3 \xi^* \right)
\no \\&
+ \frac{7}{16} \left( 2 c_1 - 3 c_2 + 3 c_3\right)
+ \frac{3}{16} e^{-\phi/2} b_z \xi^\dagger \sigma^1 \xi^*\,.
\end{align}
Taking \eqref{eq:odil}, multiplying by $\xi^\dagger \sigma^1$ and using \eqref{solb0} to eliminate $b_0$ gives
\begin{align}
\label{const2}
0 =& - \frac{1}{4} F_{0} e^{5\phi/4} \xi^\dagger \sigma^3 \xi + \frac{3}{4} \left( 2 c_1 - 3 c_2 + 3 c_3 \right) - \frac{e^{\phi/4}}{8} \left( g_z \xi^\dagger \xi^* + i h_z \xi^\dagger \sigma^3 \xi^* \right)
\cr& - \frac{e^{-\phi/2}}{4} b_z \xi^\dagger \sigma^1 \xi^*\,.
\end{align}
Adding $4/3$ of the first equation to the second equation then gives
\begin{align}
\label{algconsst}
c_1 - 2 c_2 + 2 c_3 = 0\, .
\end{align}
Finally, we take \eqref{const2} for the last algebraic equation.  This constraint is similar to the one encountered in \cite{D'Hoker:2008wc} for M-theory, where the values of $c_i$ controlled the $D(2,1;\gamma;1)$ group parameter $\gamma$.

\bibliographystyle{JHEP}
\bibliography{draft}

\end{document}